\addtocontents{}{}
\documentclass[11pt]{article}
\usepackage{amsmath,amssymb,color,graphics,epsfig,cite,footnote}
\usepackage{subfigure}
\usepackage{graphicx}
\usepackage{hyperref}
\usepackage{amsfonts}

\newcommand{\be}{\begin{equation}}
\newcommand{\ee}{\end{equation}}
\newcommand{\bea}{\setlength\arraycolsep{2pt} \begin{eqnarray}}
\newcommand{\eea}{\end{eqnarray}}
\newcommand{\nn}{\nonumber}

\usepackage[normalem]{ulem}
\usepackage{xcolor}

\def\ft#1#2{{\textstyle{\frac{\scriptstyle #1}{\scriptstyle #2} } }}
\def\fft#1#2{{\frac{#1}{#2}}}

\def\0{{\sst{(0)}}}
\def\1{{\sst{(1)}}}
\def\2{{\sst{(2)}}}
\def\3{{\sst{(3)}}}
\def\4{{\sst{(4)}}}
\def\5{{\sst{(5)}}}
\def\6{{\sst{(6)}}}
\def\7{{\sst{(7)}}}
\def\8{{\sst{(8)}}}
\def\sst#1{{\scriptscriptstyle #1}}

\begin{document}

	
	\begin{center}

 { \Large {\bf Light Ring behind Wormhole Throat:\\ Geodesics, Images and Shadows}}

		\vspace{20pt}
		
	{\large{Hyat Huang$^{1,2}$, Jutta Kunz$^2$, Jinbo Yang$^3$ and Cong Zhang$^4$ }}

\vspace{10pt}

{\it $^1$College of Physics and Communication Electronics, Jiangxi Normal University, Nanchang 330022, China\\
$^2$  Institute of Physics, University of Oldenburg, Postfach 2503, D-26111 Oldenburg,
Germany\\
$^3$ Department of Astronomy, School of Physics and Materials Science,
		Guangzhou University, Guangzhou 510006,China\\
$^4$ Institut f\"ur Quantengravitation, Friedrich-Alexander-Universit\"at Erlangen-N\"urnberg, Staudtstr. 7/B2, 91058 Erlangen, Germany
}

		\vspace{40pt}
		
		\underline{ABSTRACT}
	\end{center}

The geodesics of the Ellis-Bronnikov wormhole with two parameters are studied. The asymmetric wormhole has only one light ring and one innermost stable circular orbit located on one side of the wormhole throat. Consequently, certain light rays can be reflected back by the wormhole. Additionally, the same wormhole can have different appearances on both sides of the throat. We present novel images of the wormhole  with a light ring behind the throat in a scenario with an accretion disk as the light source and in a backlit wormhole scenario, which are distinct from the images of other compact objects and have the potential to be observed.

\vfill {\footnotesize 
~\\
hyat@mail.bnu.edu.cn\\
jutta.kunz@uni-oldenburg.de\\
yangjinbophy@gmail.com\\
czhang@fuw.edu.pl
}\ \ \ \

	\thispagestyle{empty}
	
	\pagebreak
	\tableofcontents
\addtocontents{toc}{\protect\setcounter{tocdepth}{2}}
\newpage
	


\section{Introduction}
The images of black holes captured by the Event Horizon Telescope (EHT) are significant achievements in astronomical observation \cite{ EventHorizonTelescope:2019dse, EventHorizonTelescope:2019ths, EventHorizonTelescope:2022wkp, EventHorizonTelescope:2022wok} and have sparked a great deal of interest in the theoretical study of the appearance of compact objects. 
These images can be used to test General Relativity and provide insight into the nature of gravity\cite{Younsi:2021dxe}. 
We expect to see more images in the future, not just related to black holes, but possibly to other compact objects as well.

Although images can provide valuable information on compact objects such as black holes, they may not always be sufficient to determine the underlying object with certainty. 
In the case of black holes, for instance, their defining characteristic is the presence of an event horizon. 
However, the event horizon itself is not directly visible and cannot be detected by using methods based on electromagnetic radiation.
Instead, the light ring (LR), which represents an unstable photon orbit, plays a significant role in the optical/infrared/radio appearance \cite{Cardoso:2021sip,Rosa:2023hfm} and therefore its properties have sparked  much interest \cite{Cunha:2017qtt,Guo:2020qwk,Guo:2021bcw,Cunha:2022gde}. 
While the unstable photon orbit is an important feature of black holes, it is not exclusive to them. Other types of compact objects, such as boson stars \cite{Rosa:2022tfv,Rosa:2022toh}, naked singularities \cite{Shaikh:2018lcc,Gyulchev:2019tvk,Gyulchev:2020cvo,Joshi:2020tlq, Dey:2020bgo}, quantum black holes \cite{ashtekar2018quantum,Zhang:2021xoa,Peng:2020wun,husain2022quantum,lewandowski2022quantum} and wormholes \cite{Muller:2008zza,Bambi:2013nla,Guerrero:2021pxt,Schee:2021pdt,DeFalco:2021klh,Guerrero:2022qkh,Delijski:2022jjj,Olmo:2023lil}, can also possess unstable photon orbits. 
Therefore, their visual appearance may be similar to that of black holes.
In this work, we will focus on the appearance of wormholes, to investigate its similarities and differences with respect to the images of black holes, which may help distinguish between the two types of objects and potentially shed light on the nature of spacetime itself.

Just as the horizon is important for black holes, the throat is a crucial concept for wormholes. While the throat of a static wormhole is well-defined as a minimal hypersurface of spacetime, the same cannot be said for dynamic wormholes. 
Using the trapping horizon as the throat of the dynamic wormholes is one suggestion \cite{Yang:2021diz},  but there is currently no universally accepted definition \cite{Hayward:1998pp, McNutt:2021esy, Hochberg:1998ii, Tomikawa:2015swa }. 
Defining the throat of a wormhole is essential to understand its geometry and properties. 
As shown in \cite{Nedkova:2013msa,Gyulchev:2018fmd}, the presence of a wormhole throat can modify the shape of the shadow.
In our work, focusing only on static wormholes allows for a clear definition of the throat, which can then be used to study the wormhole's visual appearance and compare it to that of black holes.

A wormhole connects two regions of spacetime through its throat. 
This may result in a special phenomenon: \textit{a light ring (LR) is located only on one side of the wormhole throat and there is no LR on the other side of the throat.} 
This means that even the same wormhole can produce different optical/infrared/radio images on the two sides of the throat. 
Indeed, as shown below, in such a case, while the wormhole can appear to be a black hole only when observed from the side of the LR, this is not the case on the other side. 
Consequently, the appearance of the wormhole viewed from the other side can be distinguished from that of black holes, boson stars, or naked singularities. 
\textit{The ability of one compact object to create two distinct optical/infrared/radio images is a typical} feature of wormholes.

Previous studies, such as \cite{Wang:2020emr,Peng:2021osd,Tsukamoto:2021fpp}, have investigated similar phenomena in thin-shell wormhole models. 
However, in this work, we will use the Ellis-Bronnikov (EB) wormhole  as an example. 
The EB wormhole, as a unique solution for a regular, static, traversable wormhole in General Relativity coupled to a free phantom scalar field,  has two asymptotically flat regions \cite{Ellis:1979bh,Bronnikov:1973fh, Yazadjiev:2017twg}. 
It is described by two parameters $(q,m)$. 
In the massless case with $m=0$, it becomes a symmetric wormhole where the LR coincides with the throat. 
Its appearance when surrounded by optically and geometrically thin dust and its lensing effects were studied in~\cite{Ohgami:2015nra,Perlick:2015vta,Bugaev:2021dna,Cai:2023ite}. For $m\neq 0$,it becomes an asymmetric wormhole, and has its LR located on one side of the throat. 
By adjusting the parameters $(q,m)$, one can make its image resemble that of a Schwarzschild black hole on the side with the LR, whereas the image on the other side looks distinctly different. 
For the case where we consider a thin accretion disk as the light source, the appearance of the wormhole on the side without the LR turns out to consist of several bright rings. 
These novel features might be observable in future astronomical observations.

The paper is organized as follows. 
In Section 2, we provide a brief overview of the EB wormhole. 
In Section 3, we review the geodesics of the EB wormhole, considering both null geodesics (i.e., for light rays) and time-like geodesics (i.e., for massive particles). 
For null geodesics, special trajectories are identified that allow light rays from one side of the wormhole to be ``reflected" back to the same side.  
Particularly, due to the presence of the LR on only one side, some of light rays from the side without LR can return after they pass through the throat.
For time-like geodesics, we have found that the innermost stable circular orbit (ISCO) is located outside of the LR on one side of the throat, and absent on the other side, meaning that the accretion disk forms only on the side of the wormhole throat with the LR.
In Section 4, we study the wormhole images in the case of an accretion disk as the light source. 
On the side with the LR, the wormhole appearance is similar to that of a Schwarzschild black hole, i.e., an image with a regular shadow and bright rings. 
On the side without the LR, the image comprises several bright rings, caused by the LR and the accretion disk on the other side. 
In Section 5, we study the backlit wormhole case by exploring three possible situations depending on the location of the screen lighting the wormhole and the observers with respect to the throat, with the assumption that the lighting screen is put far away from the throat.
Finally, in Section 6, we conclude and discuss further directions of research.
The Appendix contains some technical explanations concerning our calculations of the bright rings in the images.

\section{Wormhole geometry}

\begin{figure}[t]
\centering
\includegraphics[width=0.45\textwidth]{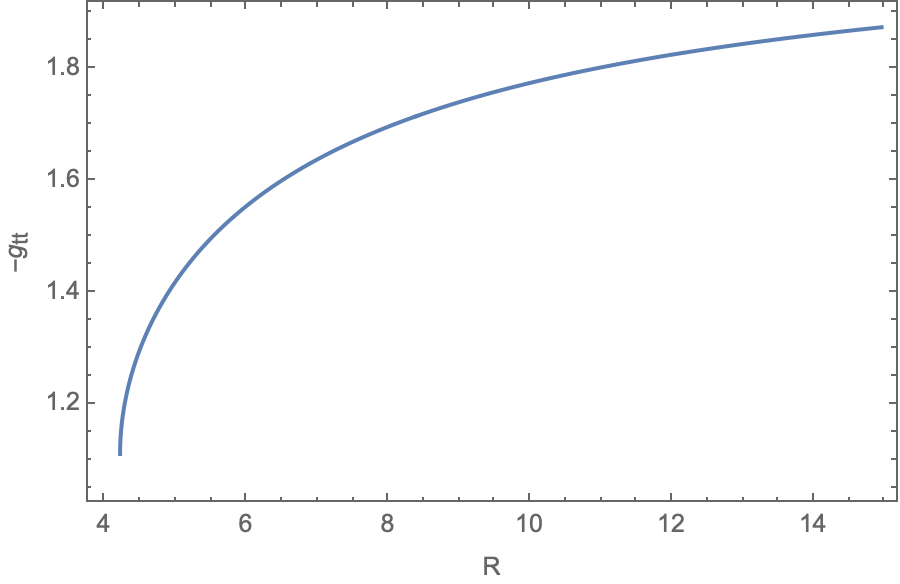} 
\includegraphics[width=0.45\textwidth]{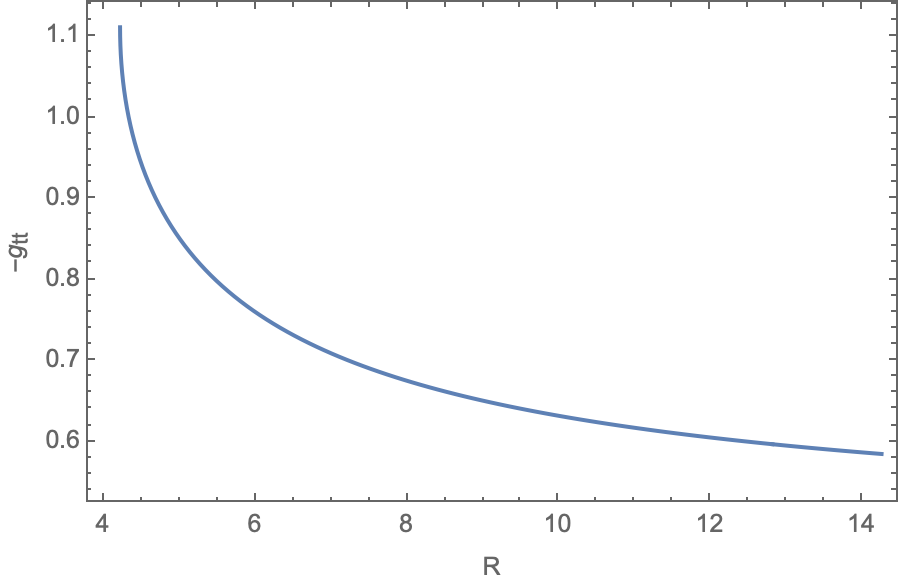} 
\caption{ \it The metric function $-g_{tt}$ as a function of the circumferential radius $R$. The left figure shows $r\to -\infty$ (Universe I), the right one $r\to+\infty$ (Universe II). 
The gravitational potential is attractive in Universe I and repulsive in Universe II.}
\label{hR}
\end{figure}

The well-known Ellis-Bronnikov (EB) wormhole is a solution of the Einstein--free scalar theory, namely
\be
{\cal L}= \sqrt{-g}(R+\ft{1}{2}(\partial \varphi)^2),
\ee
where $R$ denotes the scalar curvature and $\varphi$ is the so-called phantom field because of the sign of its kinetic term. 
The EB solution can be expressed as \cite{Huang:2020qmn}
\bea\label{bronnikov}
&&ds^2=-h(r) dt^2+ h(r)^{-1}dr^2+R^2(r)d\Omega_{2}^2\,,\nn\\
&&h=e^{-\fft{m}{q}\varphi},\qquad R^2=\fft{r^2+q^2-m^2}{h}\,,\nn\\
&&\varphi=\fft{2q}{\sqrt{q^2-m^2}}\arctan(\fft{r}{\sqrt{q^2-m^2}})\,,
\label{ebsol}
\eea
where $(m,q)$ are two integration constants. 
Note that the solution requires $q>m$. 
The wormhole throat is defined by the minimum hypersurface of the spacetime. 
It follows that the location of wormhole throat is at
\be
r_{t}=-m.
\ee
The area of wormhole throat is given by
\be
A_{t}=4\pi R^2(-m)=4\pi q^2 \exp\bigg({-\ft{2m\arctan(m/\sqrt{q^2-m^2})}{\sqrt{q^2-m^2}}}\bigg).
\ee
For any $m\neq0$, the solution \eqref{bronnikov} describes an asymmetric wormhole. 
The asymmetry not only indicates different geometries around the wormhole throat, but also reveals the presence of two different asymptotically flat regions. 
The latter can be seen by expanding the metric function for $R\to\pm\infty$. 
This shows that the gravitational potential is attractive for $R\to+\infty$ while it is repulsive for $R \to -\infty$ (see Fig.~\ref{hR}). 
An analysis of the global structure of the EB wormhole can be found in \cite{Huang:2020qmn}. 
The entire spacetime is divided into two separate universes by the wormhole throat. 
In the following we will use \textit{Universe I} to denote $r\in (-\infty, r_{t})$ and \textit{Universe II} to denote $r\in (r_t, \infty)$.

When setting $m=0$, the metric \eqref{bronnikov} reduces to the symmetric EB wormhole
\be
ds^2=-dt^2+dr^2+(r^2+q^2)d\Omega^2_2.
\ee

\section{Geodesics: light ring and ISCO}

\subsection{Geodesics}

The geodesics of a particle with energy $E$ and angular momentum $L$ in the wormhole background \eqref{bronnikov} yield for the orbits the equation
\begin{equation}
\label{nullgeo}
\begin{aligned}
\left(\frac{d r}{d\phi}\right)^2=R^4(r)\left(\frac{1}{b^2}-V(r)\right),\qquad V(r)=\frac{h(r)}{R^2(r)}-\epsilon\frac{h(r)}{L^2},
\end{aligned}
\end{equation}
where $b=\ft{L}{E}$ is the impact parameter, and $\epsilon=0$ for massless particles (photons), while $\epsilon=-1$ for massive particles.

\subsection{Light Ring}

In spherically symmetric backgrounds, the light ring is the projection of the photon sphere on the equatorial plane. 
In the EB wormhole spacetime \eqref{bronnikov}, there is only one light ring located at $r_p=-2m$.
Thus Universe I possesses a light ring. 
The wormhole mass from the view of the observers at infinity in Universe I is given by $M=m e^{\pi m/(2\sqrt{q^2-m^2})}$ \cite{Huang:2020qmn}. 
We choose $q=4.441$ and $m=1$ for all the plots throughout this paper, in order to have the location of the LR at $3$ times the wormhole mass, analogous to the Schwarzschild case.
Universe II has no light ring and only features at its inner boundary the wormhole throat. That is to say that from the view of Universe II \textit{the light ring is hidden behind the wormhole throat.}
\begin{figure}[h]
\centering
\includegraphics[width=0.5\textwidth]{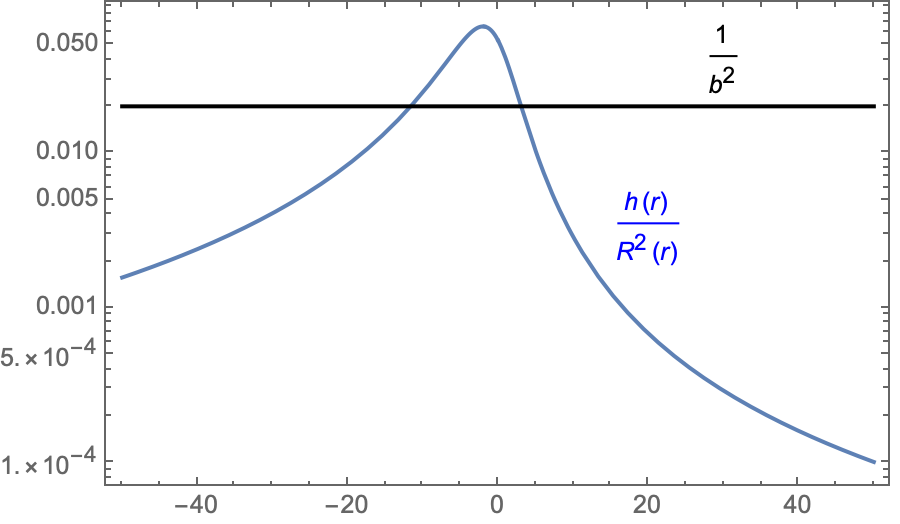}
\caption{ \it An example of $h/R^2$ and $1/b^2$ (eq.~(\ref{nullgeo})). Null geodesics are allowed only in the regions satisfying $1/b^2>h(r)/R^2(r)$. Thus, geodesics can possess a turning point.}
\label{fig:hr}
\end{figure}

\begin{figure}[t]
\centering
\includegraphics[width=0.3\textwidth]{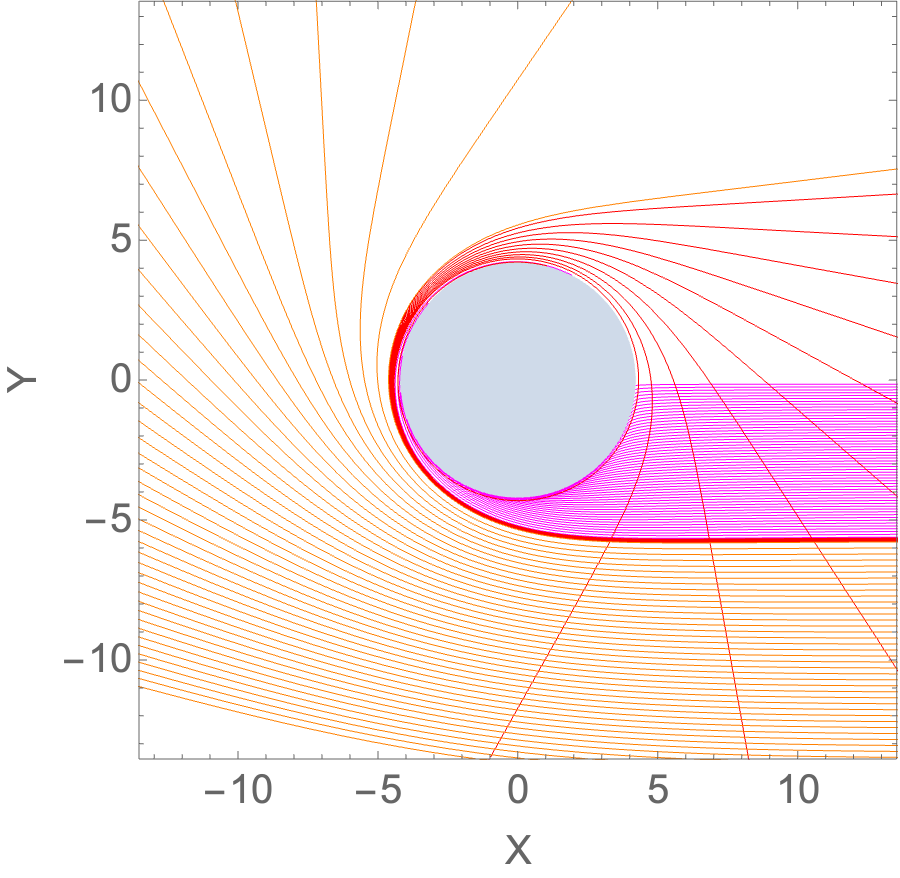} 
\includegraphics[width=0.3\textwidth]{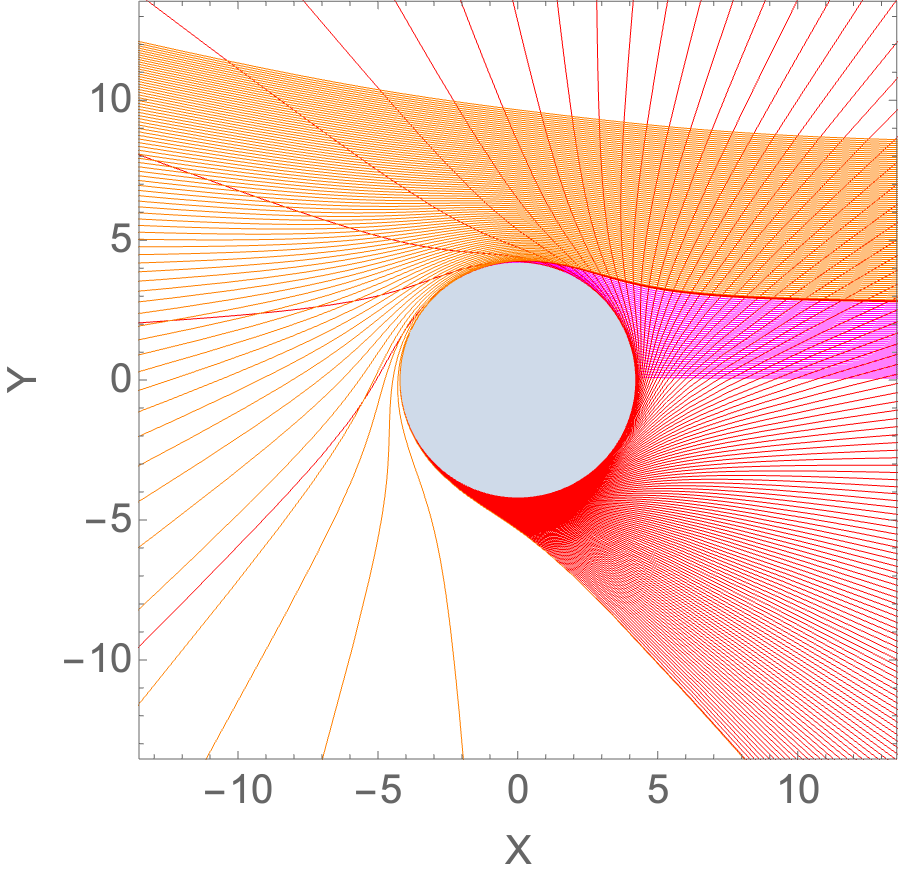} 
\includegraphics[width=0.3\textwidth]{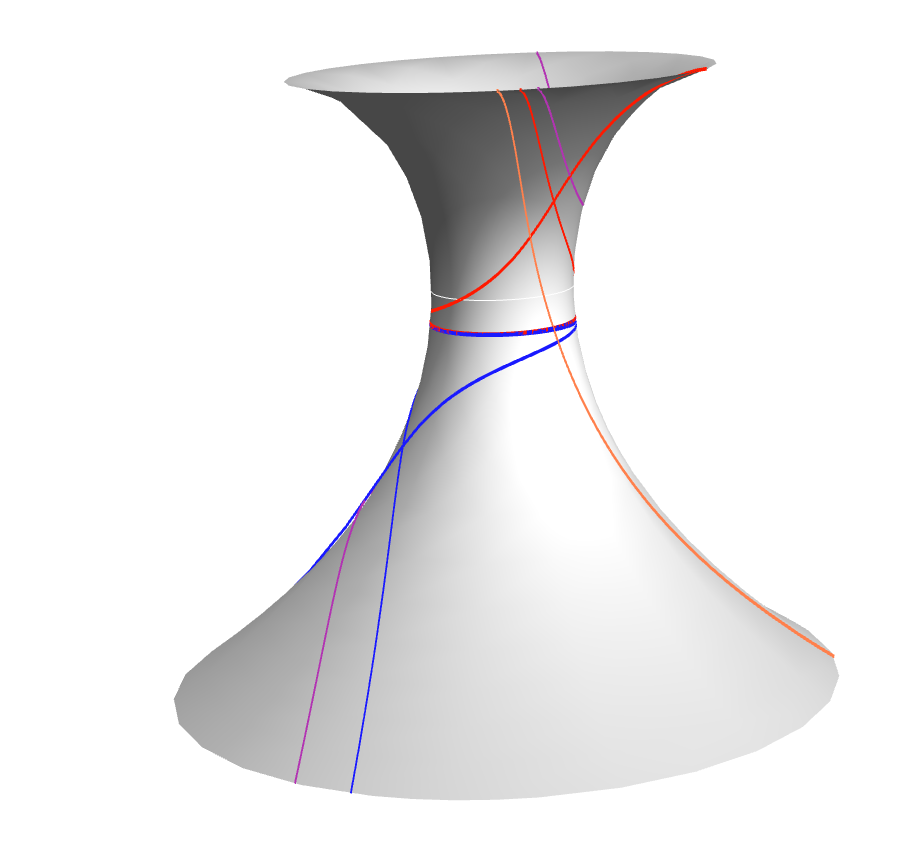} 
\caption{ \it The null geodesics of the spacetime: on the left is the figure for Universe I and in the middle for Universe II. 
Here light rays with $b>b_r$ are depicted in orange, light rays with $b_r\geq b>b_c$ in red, and light rays with $b<b_c$ in purple. 
The coordinates here are given by $\big(X=R \cos(\phi), Y=R\sin(\phi)\big)$. 
The figure on the right is an embedding diagram for the null geodesics with several typical light rays (see text). 
The upper asymptotic region represents Universe II and the lower one Universe I. 
The white circle shows the wormhole throat.}
\label{geo}
\end{figure}

To study the null geodesics, we solve \eqref{nullgeo} numerically for $\epsilon=0$.  
To get an intuitive idea before we discuss the results, it is helpful to take a look at the properties of the RHS of \eqref{nullgeo} for $\epsilon=0$. 
We illustrate these in Fig.~\ref{fig:hr}. 
For a certain value of $b$, the motion of photons is only possible in the regions where $\ft{1}{b}\geq \ft{h}{R^2}$. 
The quantity $\ft{h}{R^2}$ has a maximum at $r=-2m$, which corresponds to the location of the unstable photon orbit (light ring). 
We obtain the critical value of $b$ by solving the equation
\be
\ft{1}{b_c^2}=\ft{h(-2m)}{R^2(-2m)},
\ee
which gives
\be
b_c=e^{-\ft{2m\arctan(2m/\sqrt{q^2-m^2})}{\sqrt{q^2-m^2}}}\sqrt{3m^2+q^2}.
\ee
Therefore, there are several types of trajectories for light rays. 
We discuss these in the following for increasing values of $b$:

1) Light rays with $b<b_c$ can travel across all of the spacetime, from the asymptotic region in Universe I to the other asymptotic region in Universe II, and vice versa. 
The orange line in the embedding diagram Fig.~\ref{geo} depicts such a light ray.

2) Light rays with $b=b_c$ have a potential turning point at the light ring. 
If such a photon starts its motion in Universe I, it reaches the light ring. 
Since light rays are unstable against perturbations at the light ring, they will scatter back to infinity in Universe I as depicted by the blue line in the embedding diagram Fig.~\ref{geo} or travel across the wormhole throat to Universe II. 
Similarly, if a light ray starts its motion in Universe II it will cross the wormhole throat to Universe I and reach the light ring, from where it can either scatter back to infinity in Universe II as depicted by the red line in the embedding diagram Fig.~\ref{geo} or travel further into Universe I. 

3) Light rays with a value of $b>b_c$ represent the most interesting case. 
The figure shows two turning points, one on each side of the throat. 
In this case, a photon approaching the throat from far away in Universe I will reach the left turning point and be reflected back. 
On such a fly-by orbit it will not cross the light ring and hence will never reach the wormhole throat. 
One of the purple lines in the embedding diagram Fig.~\ref{geo} represents such a light ray.
The photons approaching the throat from Universe II, however, fall into two classes. 
For the first class the turning point is beyond the throat in Universe I.
In this case the photons will pass the wormhole throat and will then be reflected back to Universe II. 
Here the photons satisfy
\be
b_r\geq b>b_c, \qquad b_r=e^{-\ft{2m\arctan(m/\sqrt{q^2-m^2})}{\sqrt{q^2-m^2}}}q.
\ee
In contrast, for the second class the turning point is still in Universe II. 
Hence the photons are only bent by the wormhole in this case. 
One of the purple lines in the embedding diagram Fig.~\ref{geo} represents such a light ray.

Thus the following observation holds for wormhole solutions in general: 
\textit{If a wormhole spacetime contains only one light ring that does not coincide with its single throat, light rays from the side without light ring can be reflected by the throat.} 
The red line in the embedding diagram Fig.~\ref{geo} represents such a light ray.

We exhibit the null geodesics of the asymmetric wormhole spacetime in a systematic fashion in the left and middle plots of Fig.~\ref{geo}. 
The left figure shows the geodesics for Universe I and the middle figure for Universe II. 
Here the orange lines depict light rays with $b>b_r$, the red lines depict light rays with $b_r\geq b>b_c$, and the purple lines depict light rays with $b<b_c$. 
\footnote{Note, that in the left figure we do not show the light rays of the first class entering Universe I via the throat that are reflected back.}
The grey center represents the wormhole throat.

\subsection{ISCO}

\begin{figure}[t]
\centering
\includegraphics[width=0.375\textwidth]{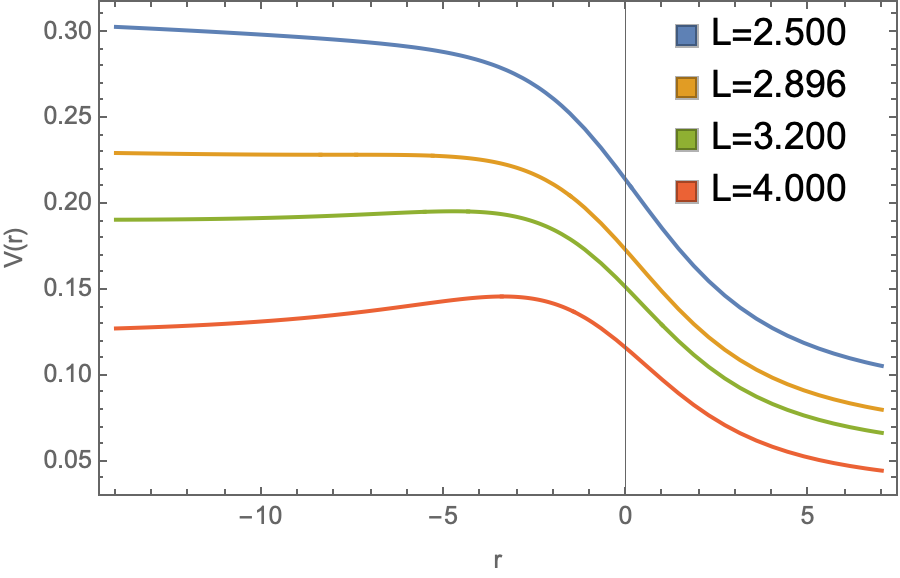} 
\includegraphics[width=0.4\textwidth]{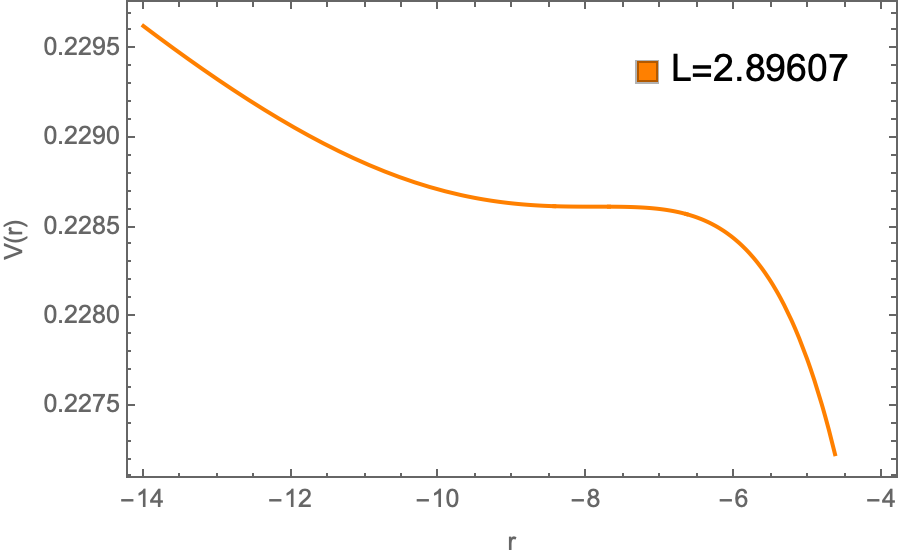} 
\caption{ \it The effective potential $V(r)$ (eq.~(\ref{nullgeo})) is shown for timelike geodesics ($\epsilon=-1$) and several values of the angular momentum $L$ in the left figure, and for the critical value $L=2.89607$ of the ISCO in the right figure.}
\label{V}
\end{figure}

The existence of bound orbits of massive particles is important in order to form an accretion disk around a compact object. 
In this case we should set $\epsilon=-1$ in \eqref{nullgeo}. 
Of particular interest here is the innermost stable circular orbit (ISCO). 
To investigate the ISCO, one needs to find the minimum value of $L$ such that 
\be
\ft{\partial V}{\partial r}|_{r=r_{ISCO}}=0, \qquad \ft{\partial^2 V}{\partial^2 r}|_{r=r_{ISCO}}=0.
\ee
When solving the above equations, we see that the wormhole spacetime possesses only one ISCO, which is located at
\be
r_{isco}= -3m-\sqrt{4m^2+q^2},
\ee
and
\be
L=-\ft{\sqrt{2m(-q^2-3m(2m+\sqrt{4m^2+q^2}))}}{\sqrt{-m-\sqrt{4m^2+q^2}}}\exp\bigg(-\ft{m\arctan(\ft{3m+\sqrt{4m^2+q^2}}{\sqrt{q^2-m^2}})}{\sqrt{q^2-m^2}}\bigg).
\ee
Thus there is no ISCO in Universe II. 
In order to get an intuitive idea of the possible orbits, we exhibit plots of $V$, eq.~(\ref{nullgeo}), in Fig~\ref{V}. 
For our choice of parameters $q=4.441,m=1$, the ISCO is obtained for $L=2.896$ and located at $r=-7.871$. 
Note, that the effective potential $V(r)$ is monotonically decreasing in Universe II, hence there are no bound orbits for massive particles. 
Based on the above considerations we draw the conclusion that only one side of the wormhole throat allows for the existence of an accretion disk. 

For the symmetric EB wormhole, where $m=0$, the single light ring coincides with the wormhole throat. 
In this case there is no stable bound orbit and no ISCO in either Universe ($L=0$ for $m\to 0$). 
Thus, in the symmetric wormhole case, no accretion disk would exist around the wormhole throat.

\section{Wormhole images with an accretion disk}\label{acc}

In this section, we present the results for wormhole images when the emission of light originates from an optically and geometrically thin disk around the wormhole throat. 
According to the above considerations, the accretion disk will be surrounding the wormhole only in Universe I.

\begin{figure}[t]
\centering
\includegraphics[width=0.4\textwidth]{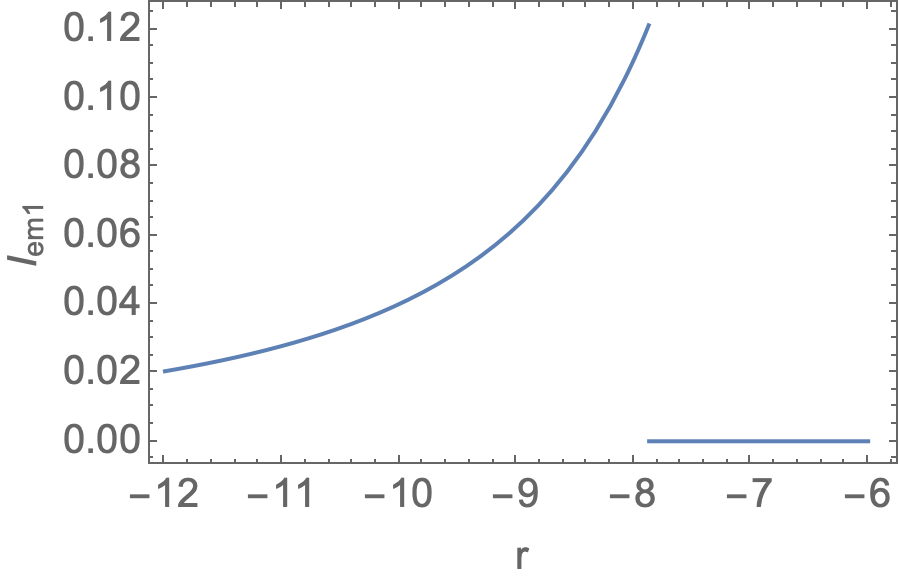} 
\includegraphics[width=0.4\textwidth]{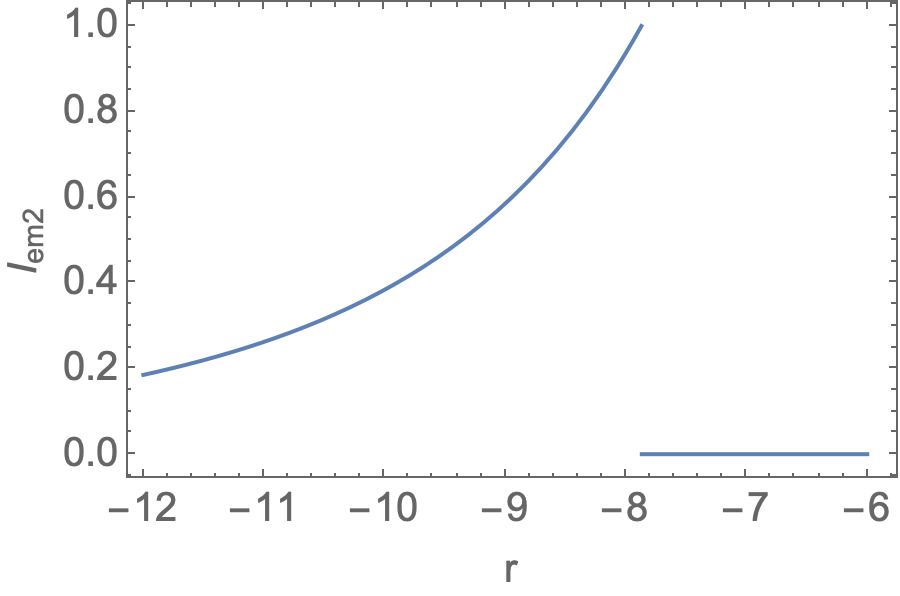} 
\caption{\it The emitted intensity $I_{em}$ (eq.~(\ref{intensity})): the left figure shows $I_{em1}$ and the right one $I_{em2}$.}
\label{em}
\end{figure}

We here adopt the following two models for the emitted intensity
\begin{eqnarray}
&&I_{em1}=
\begin{cases}
\frac{1}{(-r-5)^2}, & r\leq r_{ISCO}\\ \nn
0, &  r>r_{ISCO},
\end{cases}\\
&&I_{em2}=(\frac{r_{ISCO}}{r})^4\frac{1+\tanh(50(r-r_{ISCO}))}{2},
\label{intensity}
\end{eqnarray}
which arise in astrophysics scenarios (see e.g.~\cite{Rosa:2022tfv, Gralla:2019xty, Li:2021riw,Guerrero:2022msp, Guerrero:2022qkh, Olmo:2021piq}). 
We show these two functions employed for the emitted intensity in Fig.~\ref{em}.
\begin{figure}[h!]
\centering
\includegraphics[width=0.35\textwidth]{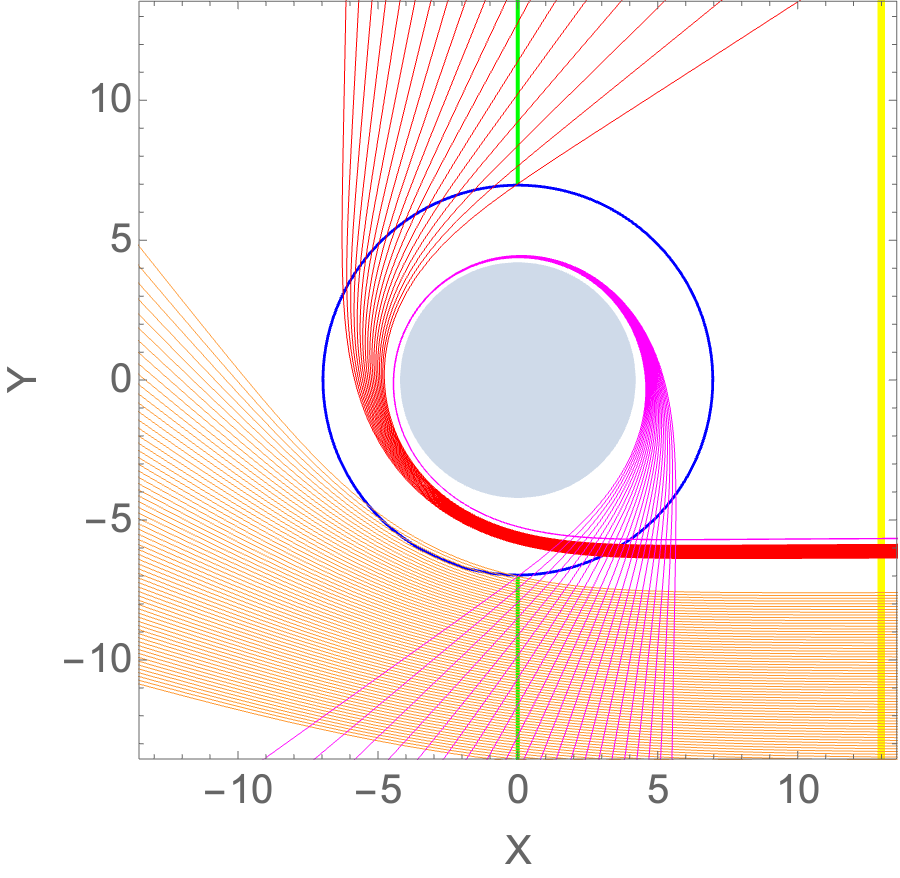}  \qquad\includegraphics[width=0.4\textwidth]{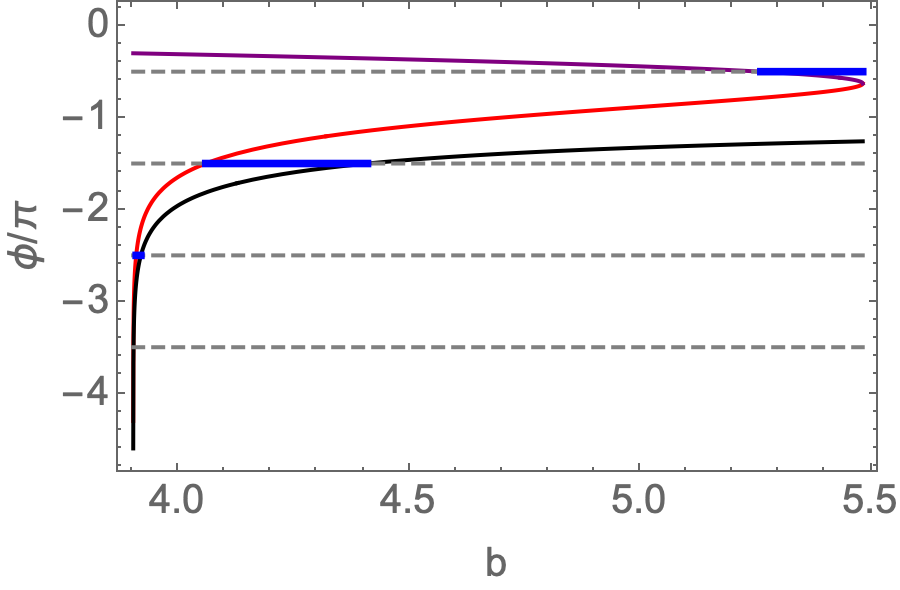}
\caption{\it The left figure shows the configuration in Universe I: The green lines indicate the accretion disk, and thus the light source. 
The blue circle denotes the location of the ISCO. 
The yellow line denotes the observation screen in the asymptotic region of Universe I.
Also shown are the sets of light rays producing the image of the wormhole on the screen (see text). 
In the right figure the black line denotes the total deflection angle of a light ray 
approaching from infinity and scattered back to infinity in Universe I.
The red line denotes the deflection angle of a light ray from infinity in Universe I to the ISCO. 
The three blue lines, from left to right, correspond to the range of the purple lines, the red lines, and the orange lines of the left figure.
As seen in the left figure, a light ray may cross the ISCO twice and will then give rise to two different deflection angles at $R=R_{isco}$. 
We show the first angle and the second angle as functions of $b$ in purple and in red, respectively, in the right figure. 
For the orange light rays of the left figure, the first intersection with the ISCO taking place at $\phi=-\ft{\pi}{2}$ corresponds to the edge of the outer bright region in Fig.~\ref{image1}. 
For the red and purple light rays of the left figure, in contrast the second intersection with the ISCO is crucial to determine the boundary of the bright regions. 
Thus, for the red and purple light rays the red line in the right figure is a boundary line for the bright regions, when its value is $-\ft{3\pi}{2}$, $-\ft{5\pi}{2}$ and so on.
The other boundary line is given by the black line, since the light rays reach the accretion disk at infinity.}
\label{ob}
\end{figure}
\begin{figure}[h!]
\centering
\includegraphics[width=0.35\textwidth]{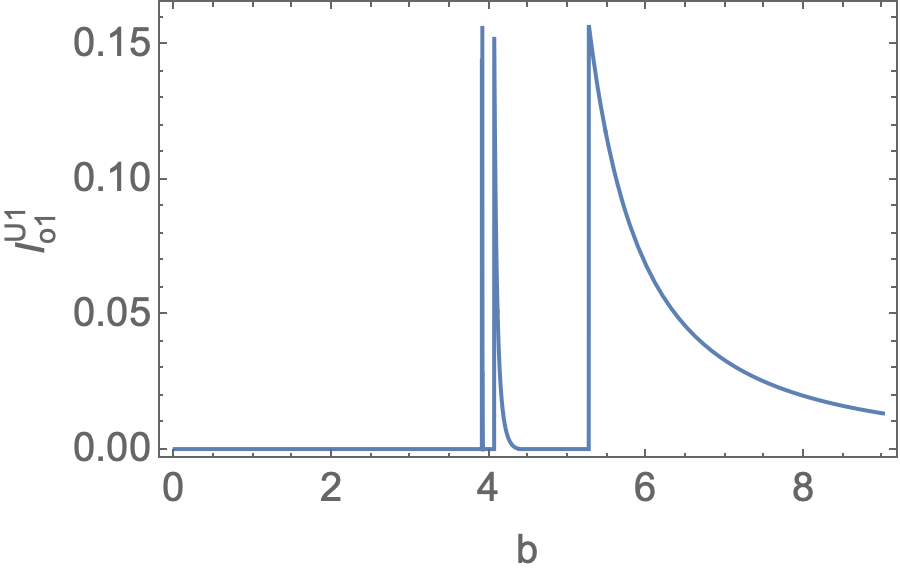} \qquad\qquad\quad
\includegraphics[width=0.30\textwidth]{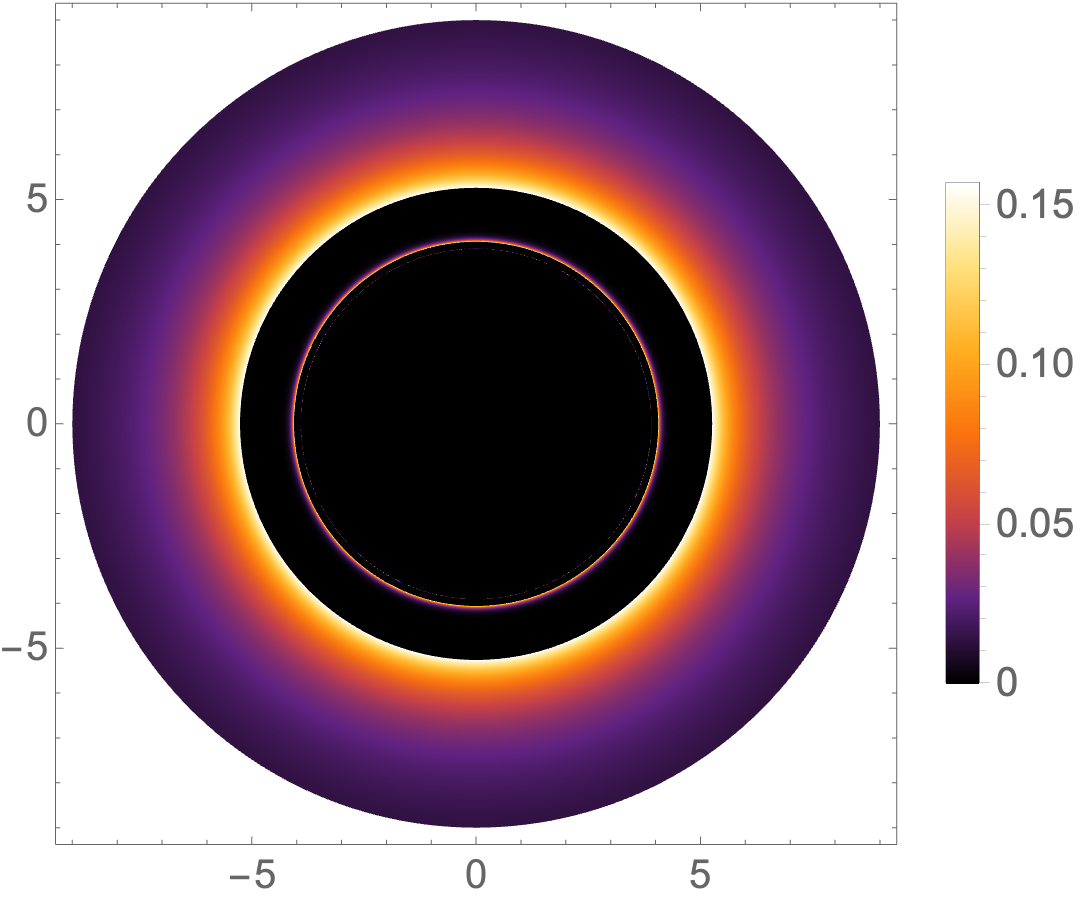} 
\includegraphics[width=0.35\textwidth]{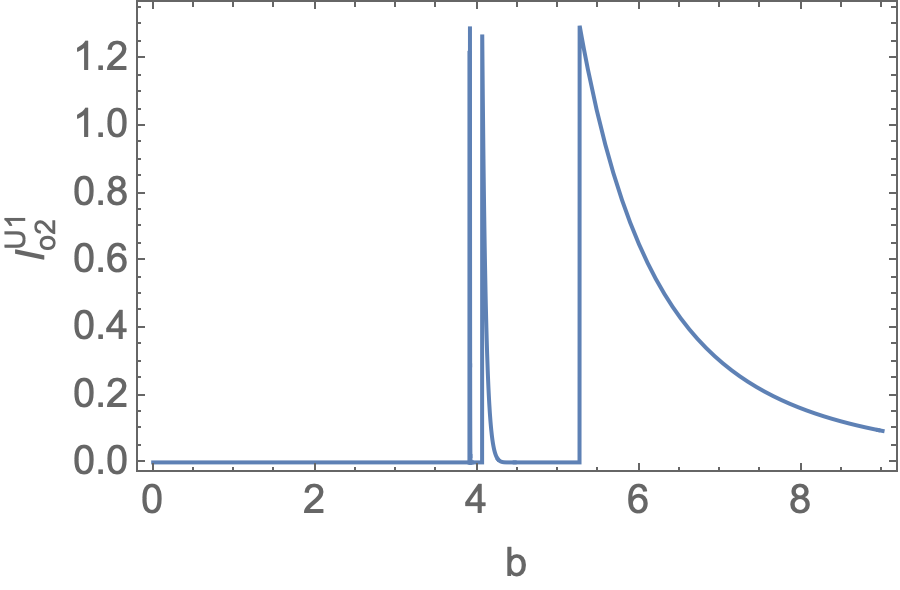} \qquad\qquad\quad
\includegraphics[width=0.30\textwidth]{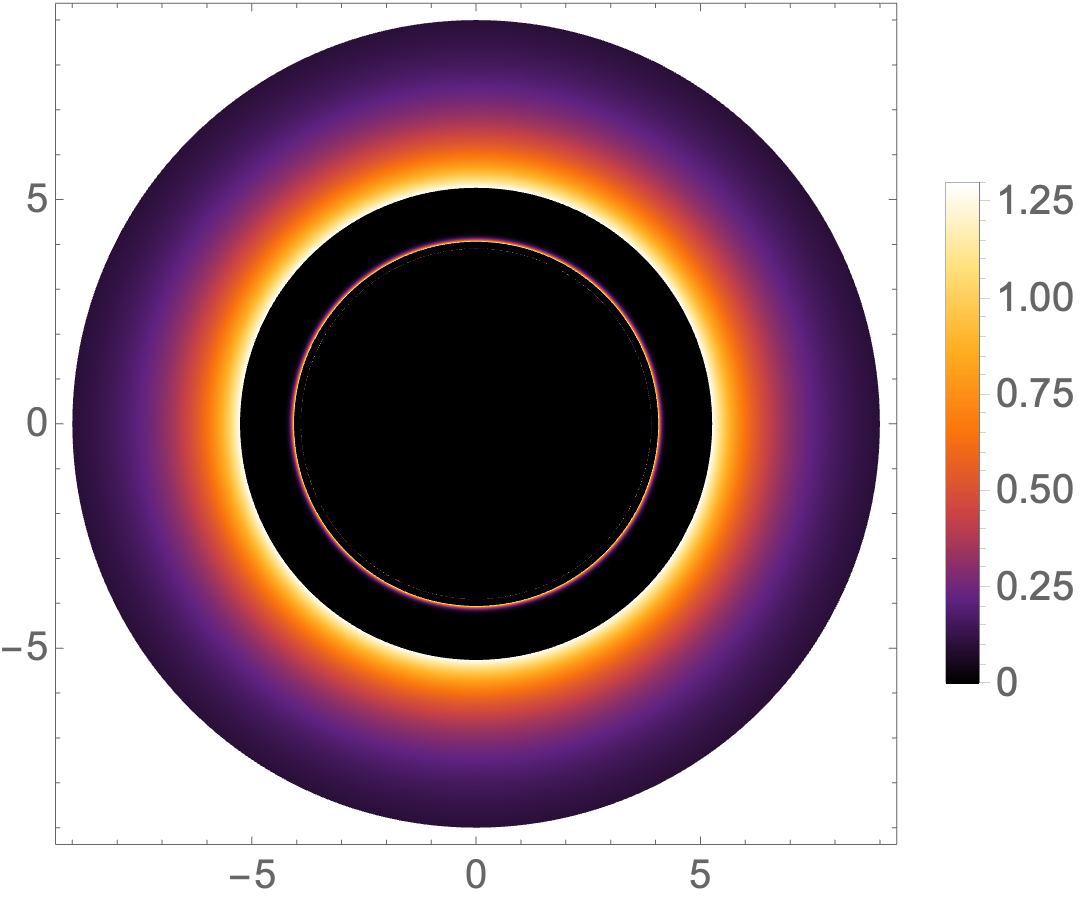} 
\caption{\it The dependence on the impact parameter $b$ is shown for the observed intensities and illustrated in the corresponding images for model $I_{em1}$ (upper figures) and model $I_{em2}$ (lower figures). 
The observers here are in Universe I.
}
\label{image1}
\end{figure}

The emitted intensity $I_{em}$ and the received intensity $I_{obs}$ have the relation \cite{Guerrero:2022msp, Guerrero:2022qkh, Olmo:2021piq}
\be\label{inten}
I_{obs}(b)= 
\sum_n \left(\frac{g_{tt(r)}}{g_{tt}(\pm\infty)} \right)^2\, I_{em}|_{r=r_n(b)} ,
\ee
where $r_n(b)$ is the transfer function that indicates the $n^{th}$ intersection with the accretion disk. 
Now we use the ray-tracing method to obtain the observed intensity $I_o$ and the wormhole images. 
One can find the details of the method in Refs.~\cite{Gralla:2019xty,Zhang:2023okw}. 
Applying the usual set-up, the resulting configuration is shown in Fig.~\ref{ob}.
Here the location of the accretion disk, i.e., the light source, is seen in green. 
It extends outward from the ISCO (blue circle).
The observation screen is located asymptotically far from the wormhole in Universe I and is shown in yellow.

To construct the image for observers at infinity in Universe I, we now apply the ray-tracing method to the set-up in Fig.~\ref{ob}.
This shows that only those light rays which intersect the accretion disk can be observed. 
For example, the light rays depicted in orange produce the outer, most bright region of the right images shown in Fig.~\ref{image1}, for which $b\in (5.2614,\infty)$. 
The light rays with $b=5.2614$ are associated with the edge of the shadow. 
The light rays depicted in red result in the first bright ring inside the shadow in the images of Fig.~\ref{image1}, and feature $b\in (4.0593,4.4097)$. 
Every time the light rays bend by $\pi$, this will cause a bright ring. So, there are in fact many bright rings inside the shadow. 
The widths of the rings are determined by the corresponding ranges of $b$. But these ranges become narrower and narrower. The light rays in purple cover the small range of $b\in (3.9087, 3.9189)$, which is therefore too hard to observe in Fig.~\ref{image1}. 
Recalling that the critical value of the LR is $b=b_c=3.9023$, we realize that the purple light rays are already very close to the LR. 
Hence, this implies that observations will yield only one bright ring  even though there are many thin bright rings inside the shadow, that get thinner and thinner.

We show the results in Fig.~\ref{image1}. 
For observers in Universe I the image of an EB wormhole looks like that of a Schwarzschild black hole as obtained in Ref.~\cite{Gralla:2019xty}. 
The shadows cast by the emission profiles are bounded by the ISCO, where the critical light ray has $b=5.2614$. 
Moreover, there are bright rings inside the black hole shadow, close to the LR.

\begin{figure}[t]
\centering
\includegraphics[width=0.35\textwidth]{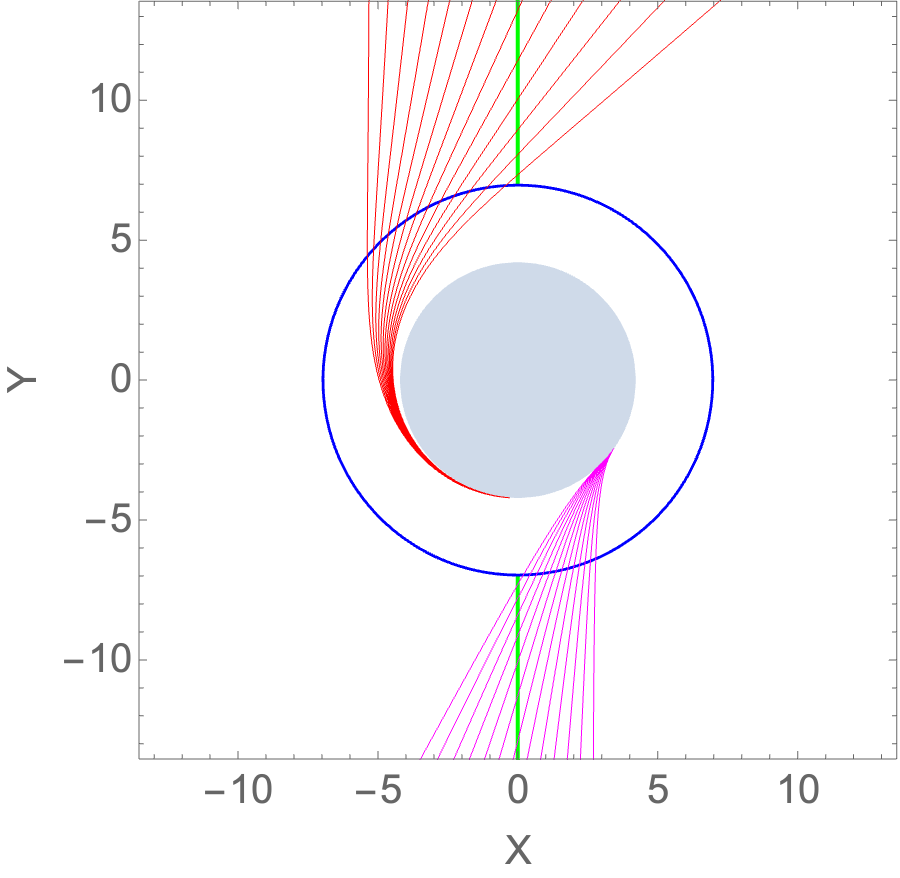}  \qquad\includegraphics[width=0.4\textwidth]{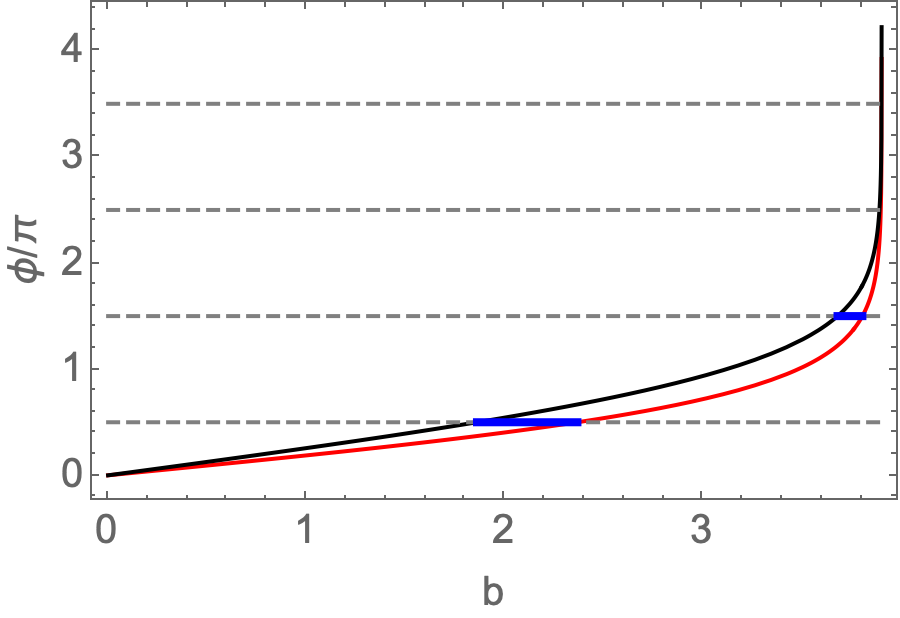}
\caption{\it The left figure shows light rays from infinity in Universe II that travel across the throat to Universe I, and there hit the accretion disk. 
The range of the red lines corresponds to the first left blue line in the right figure. 
The range of the purple lines corresponds to the second blue line in the right figure. 
In the right figure the red line denotes the deflection angle for light rays from infinity in Universe II to the ISCO in Universe I. 
The black line denotes the deflection angle for light rays from infinity in Universe II to infinity in Universe I.}
\label{wd}
\end{figure}

\begin{figure}[t]
\centering
\includegraphics[width=0.35\textwidth]{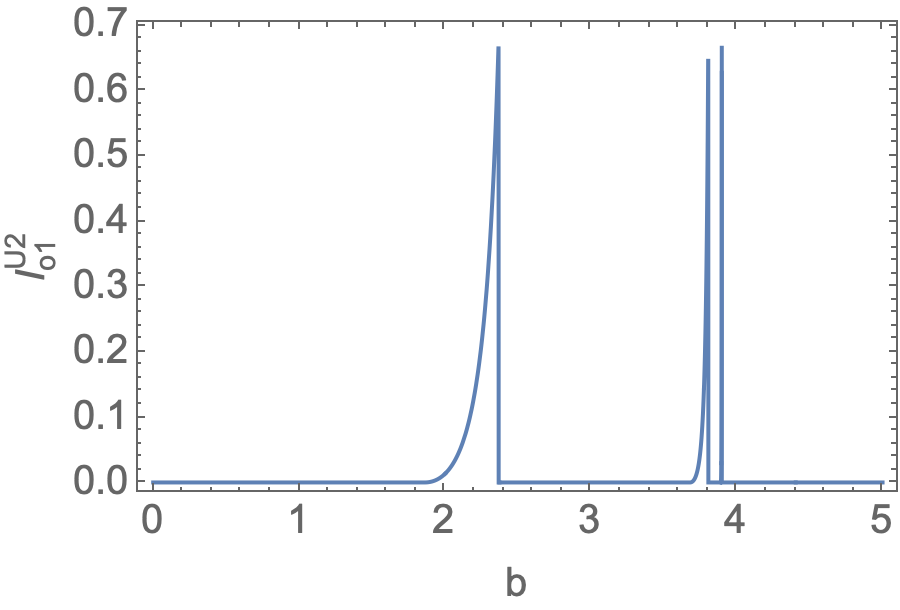} \qquad\qquad\quad
\includegraphics[width=0.30\textwidth]{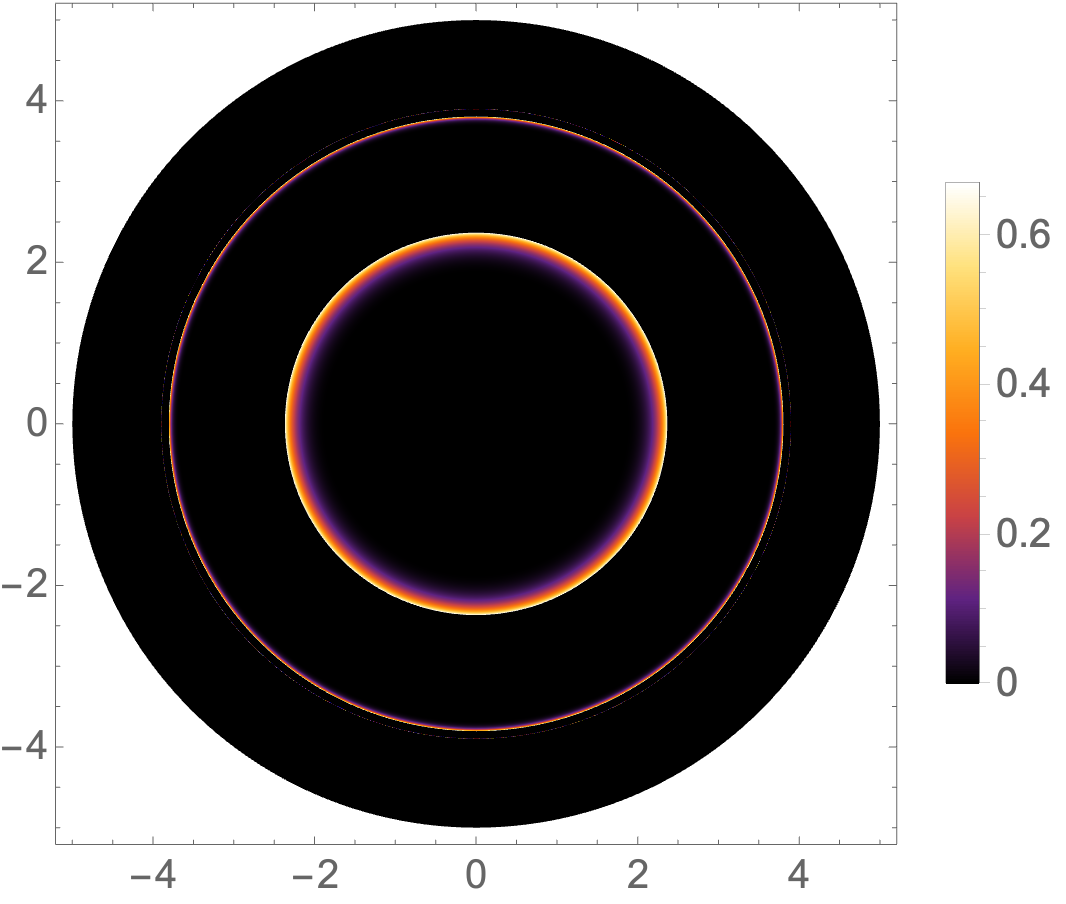} 
\includegraphics[width=0.35\textwidth]{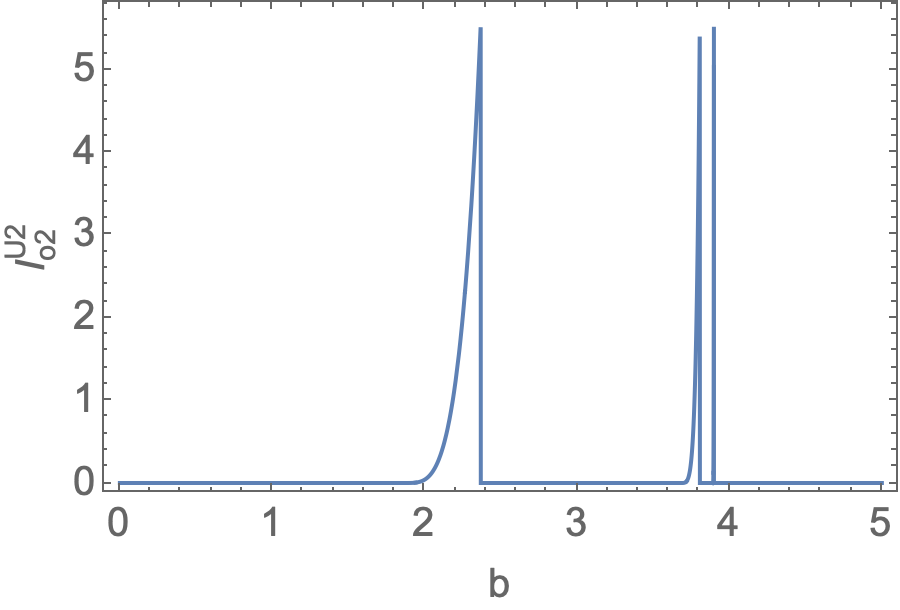} \qquad\qquad\quad
\includegraphics[width=0.30\textwidth]{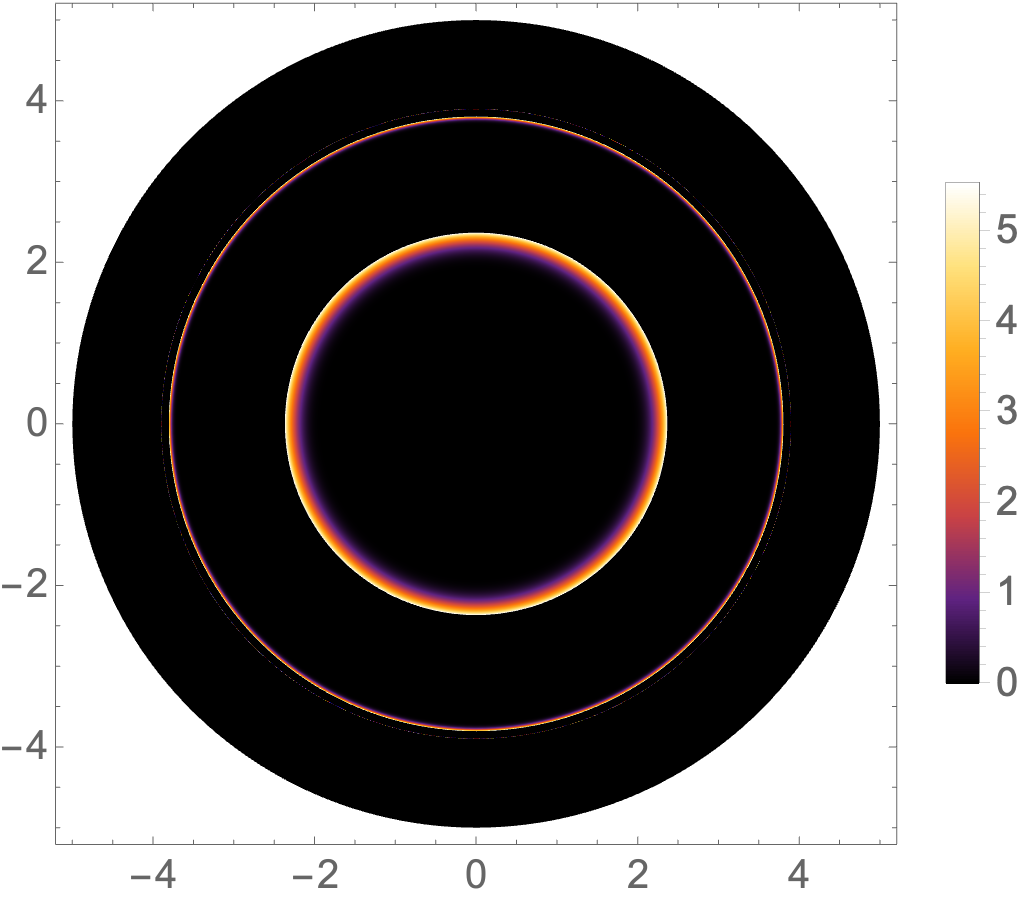}
\caption{\it The dependence on the impact parameter $b$ is shown for the observed intensities and illustrated in the corresponding images for model $I_{em1}$ (upper figures) and model $I_{em2}$ (lower figures). 
The observers here are in Universe II.
}
\label{image2}
\end{figure}

For observers at infinity in Universe II, there is no light source and all the light originates from Universe I. 
Light rays that can cross the wormhole throat satisfy $b<b_c$. 
But not all these light rays can intersect the accretion disk in Universe I. 
To clarify this issue, we show the related light rays and the deflection angle of the light rays from $+\infty$ to $-\infty$ (black line) and from $+\infty$ to $r_{isco}$ (red line) and their variation with $b$ in blue in Fig.~\ref{wd}. 
The corresponding images are shown in Fig.~\ref{image2}. 

There is no accretion disk or other light source in Universe II, therefore we here observe only light rays from Universe I. 
The innermost bright ring is produced by light rays with $b\in(1.8614, 2.3679)$ (purple lines in Fig.~\ref{wd}). 
The second innermost bright ring is produced by light rays with $b\in (3.6799,3.8052)$ (red lines in Fig.~\ref{wd}). 
Since light rays can wind many times for $b$ close to $b_c$, we conclude that there are in fact many bright rings. 
However, the width of these bright rings decreases and so does the distance between two such bright rings.
Thus, we can not observe these further bright rings. 
This intriguing phenomenon is new in astrophysics.
It indicates that the image of a spherically symmetric wormhole could consist of a multitude of rings with decreasing width and increasing size, that approach the LR at $b_c=3.9023$.

\section{Backlit wormhole}\label{back}

We next discuss the appearance of the backlit wormhole, where we assume a planar screen that is infinitely far away, and infinite in extent, and that emits isotropically with uniform brightness behind a wormhole. 

In Ref.~\cite{Gralla:2019xty} this model was considered for the Schwarzschild black hole. 
We here follow their conventions and assume that the observed brightness $I_{ob}$ is equal to the emitted brightness $I_{em}$. 
But otherwise the situation for a wormhole is different. 
In a wormhole spacetime, there can be two screens that lie in two different Universes, respectively. 
This fact can lead to new interesting phenomena concerning the wormhole appearance.

\begin{figure}[t]
\centering
\includegraphics[width=0.4\textwidth]{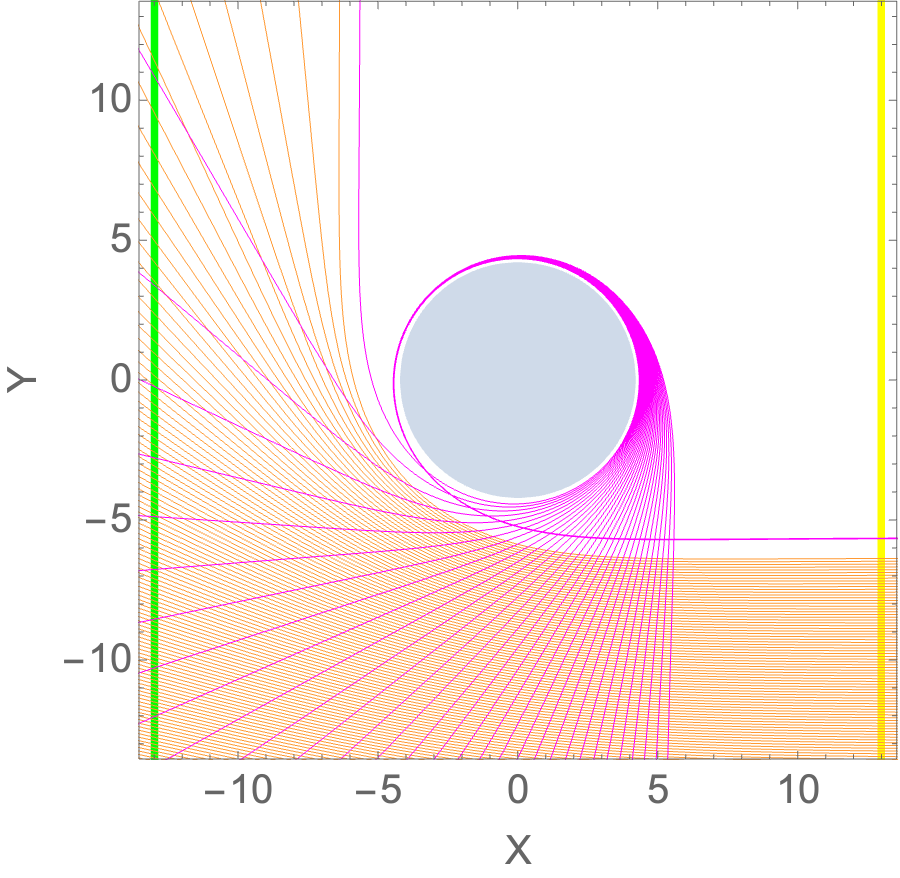} 
\caption{\it The figure shows Universe I. 
The green line denotes the emitting screen and the yellow line the observation screen in the asymptotic region of Universe I. 
The orange lines represent light rays that hit the emitting screen with a deflection angle in $(1/2 \pi, \pi)$, while the red lines depict light rays that hit the emitting screen with deflection angle in $(3/2 \pi, 2\pi)$.}
\label{backlight}
\end{figure}

\begin{figure}[h!]
\centering
\includegraphics[width=0.3\textwidth]{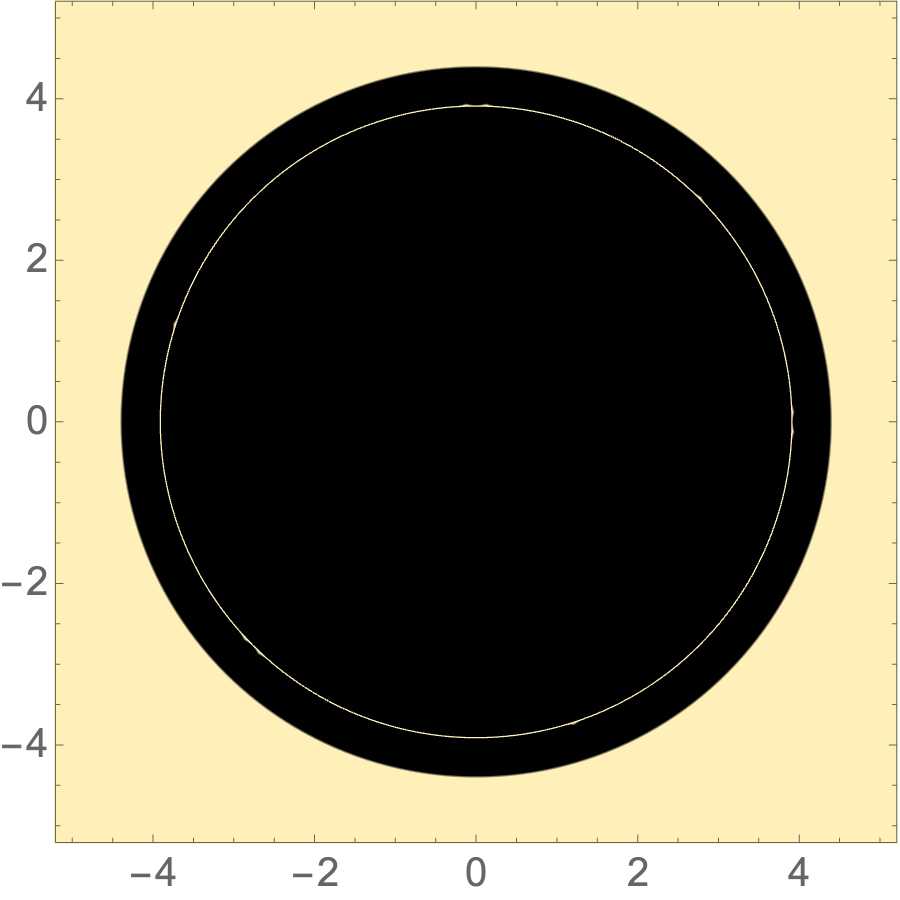} 
\includegraphics[width=0.3\textwidth]{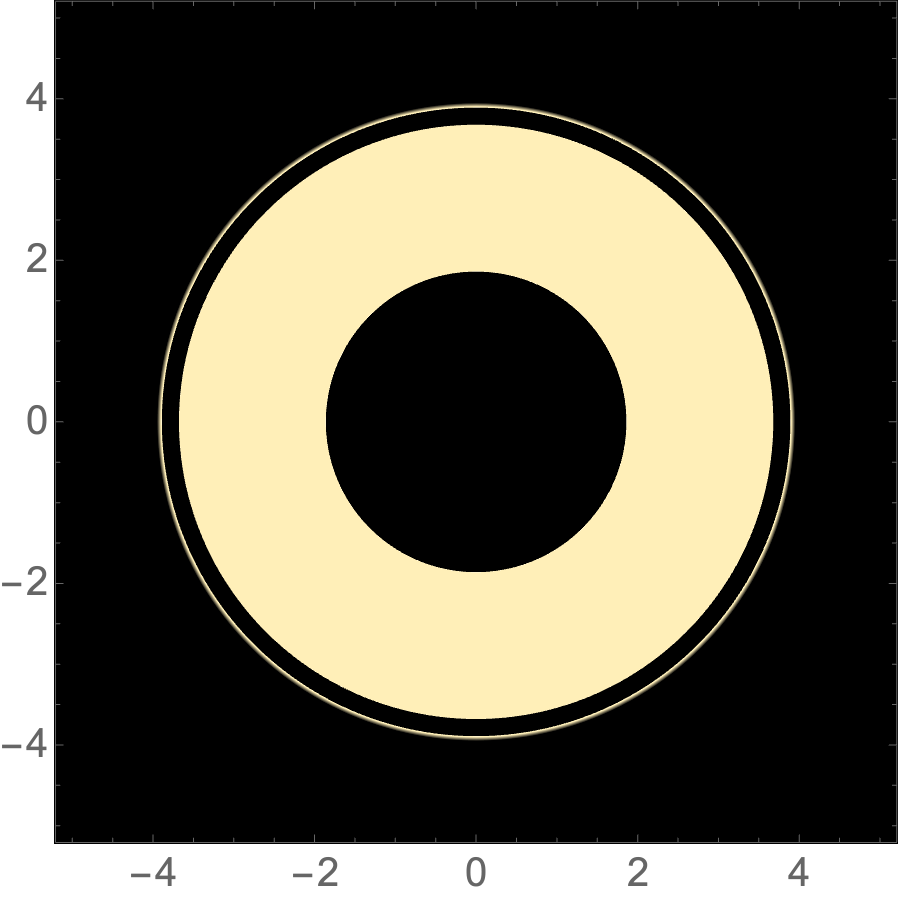} 
\includegraphics[width=0.3\textwidth]{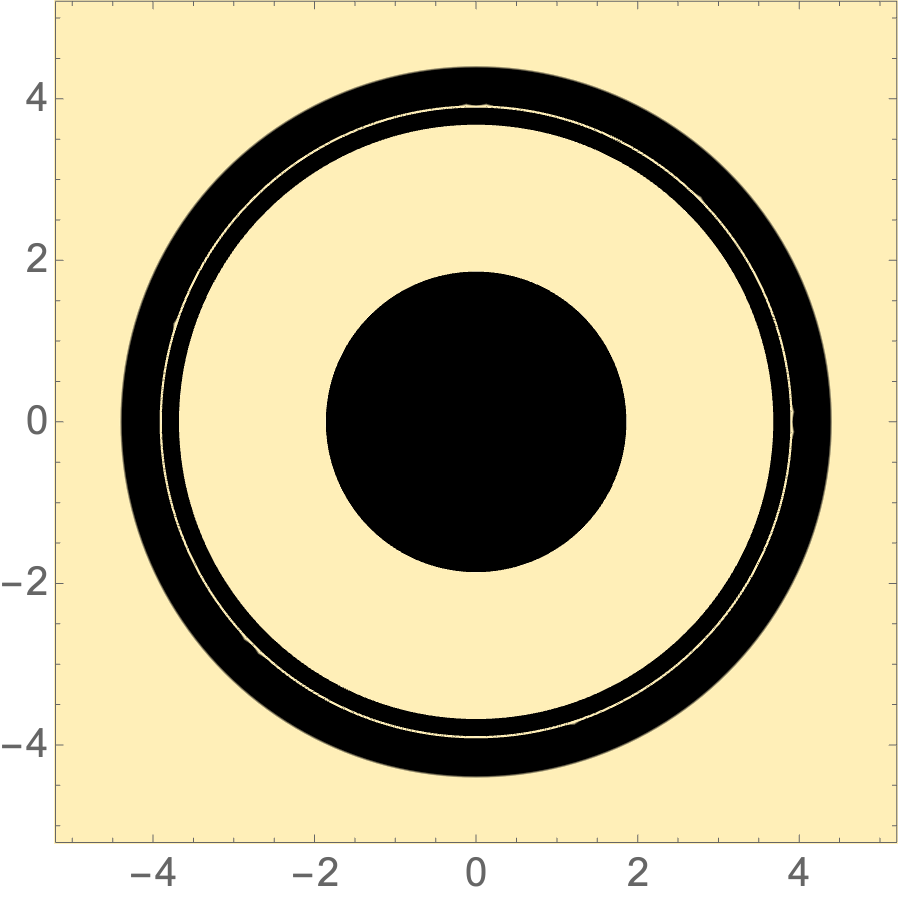} 
\caption{ \it The figures show the backlit wormhole images. 
In the left figure the emitting screen and the observers are both in Universe I. 
In the central figure the emitting screen and observers are in different Universes. 
In the right figure emitting screens are in both Universes but the observers are in Universe I.
} 
\label{back1}
\end{figure}

To begin with, let us consider that the emitting screen is located in Universe I, and the observers are also in Universe I (see Fig.~\ref{backlight}). 
The corresponding wormhole image is shown in the left plot of Fig.~\ref{back1}. 
The edge of the shadow pertains to the edge of the orange lines in Fig.~\ref{backlight}, where $b=4.4097$. 
The cause of the bright ring inside the shadow are the light rays with $b\in (3.9031, 3.9189)$ with deflection angle in $(3/2 \pi, 2\pi)$.
Similar to the case of the accretion disk, there are in fact many bright rings inside the shadow due to the multiple winding of light rays close to the LR. 
But this is hard to observe.

\begin{figure}[h!]
\centering
\includegraphics[width=0.4\textwidth]{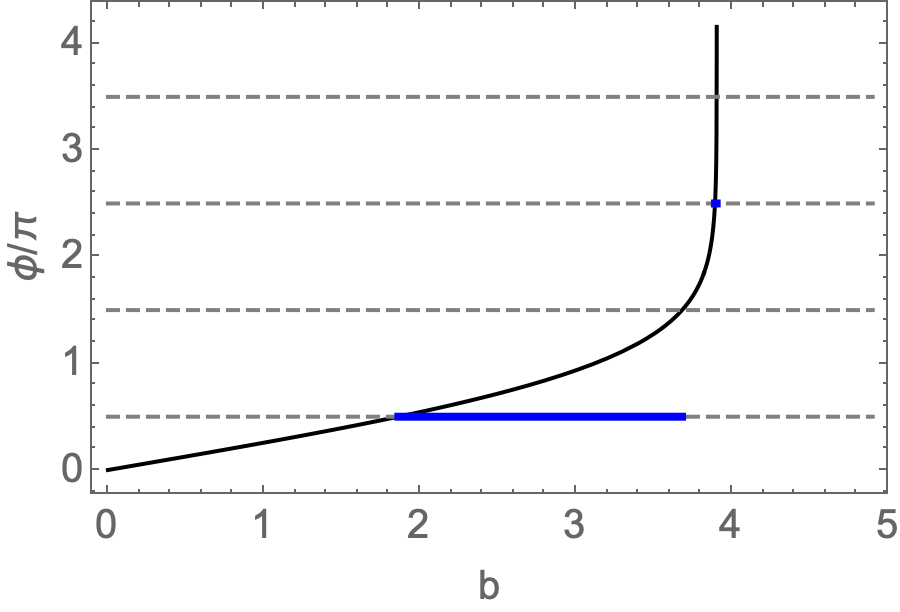} 
\caption{\it In the figure the black line denotes the deflection angle for light rays from infinity in Universe II to infinity in  Universe I. 
The long blue line corresponds to the bright region with $b\in(1.8614,3.6799)$ in the central plot of Fig.~\ref{back1}. 
The short blue line corresponds to the thin bright ring with $b\in (3.89116,3.9019)$.}
\label{wd2}
\end{figure}

Next, let us consider that the screen and the observers are not in the same Universe. 
One may naively think that the appearance of the wormhole is different when the screen is in Universe I and when it is in Universe II. 
However, this is wrong. 
No matter in which Universe the screen is, the observers in the other Universe obtain the same image. 
We show the image for the case with the screen in Universe II and the observers in Universe I in the central plot of Fig.~\ref{back1}. 
Note that only light rays with $b>b_c$ can then travel across the wormhole throat from one asymptotic infinity to the other asymptotic infinity. 
Moreover, not all such light rays can hit the emitting screen behind the wormhole throat. 
Only light rays with $b<b_c$ for which the deflection angle satisfies $\Delta \phi \in (1/2\pi+n \pi, 3/2\pi+n \pi)$ with integer $n$ can cause bright regions. 
The ranges of the corresponding light rays are shown in blue in Fig.~\ref{wd2}. 
The bright region in the central plot of Fig.\ref{back1} is caused by light rays with $b\in (1.8614,3.6799)$. 
The thin bright ring outside the bright region corresponds to light rays with $b\in (3.89116,3.9019)$. 
Further bright rings are hard to observe.

Finally, if the emitting screens are in both Universes and possess the same brightness, then observers in Universe I see the right image shown in Fig.~\ref{back1}, which depicts a brought bright ring surrounded by a narrow black ring, and a shadow in the center. 
In fact, combining the left and the central plot of Fig.~\ref{back1} gives rise to the right plot of Fig.~\ref{back1}.

\begin{figure}[h!]
\centering
\includegraphics[width=0.4\textwidth]{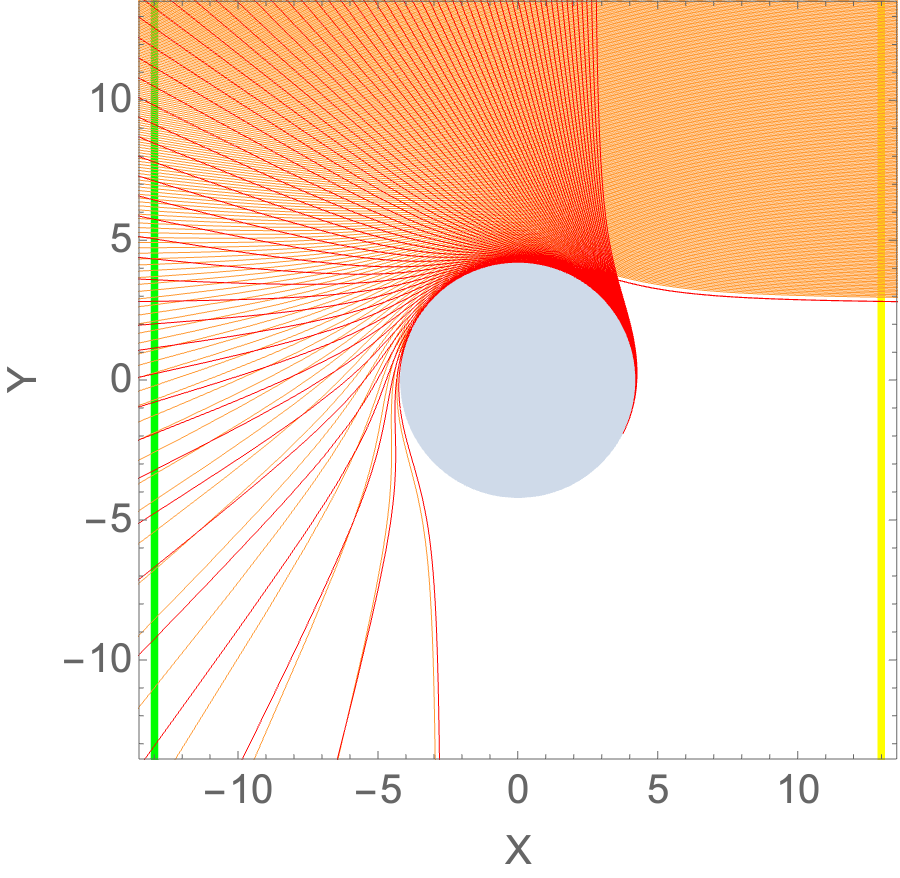} 
\caption{\it The figure shows Universe II. 
The green line denotes the emitting screen, and the yellow line the observation screen in the asymptotic region of Universe II. 
The orange lines represent light rays that hit the emitting screen with a deflection angle in $(1/2 \pi, \pi)$, while the red lines depict the light rays that hit the emitting screen with deflection angle in $(3/2 \pi, 2\pi)$.}
\label{backlight2}
\end{figure}

\begin{figure}[h!]
\centering
\includegraphics[width=0.3\textwidth]{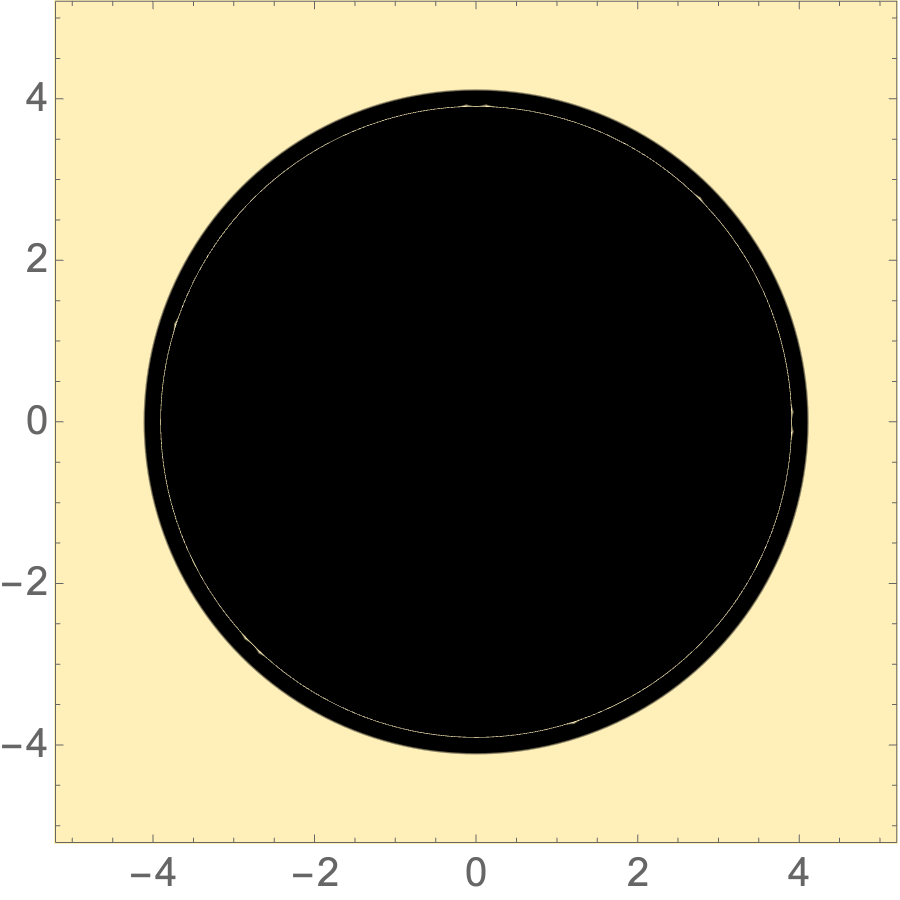} 
\includegraphics[width=0.3\textwidth]{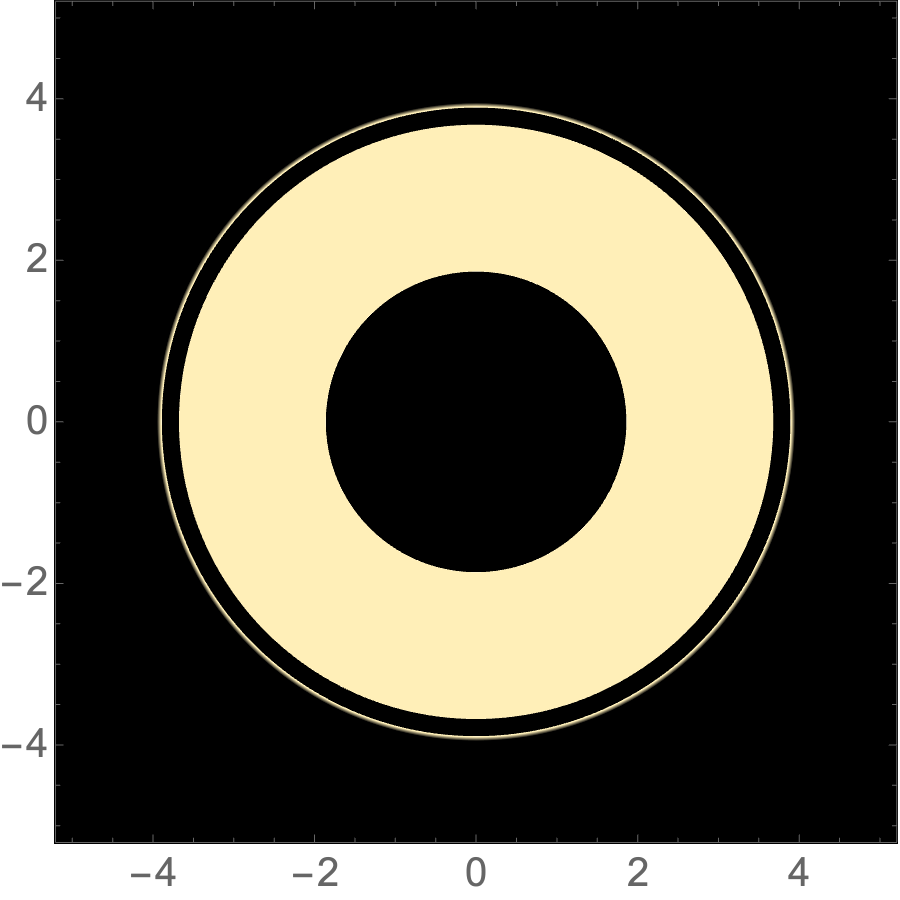} 
\includegraphics[width=0.3\textwidth]{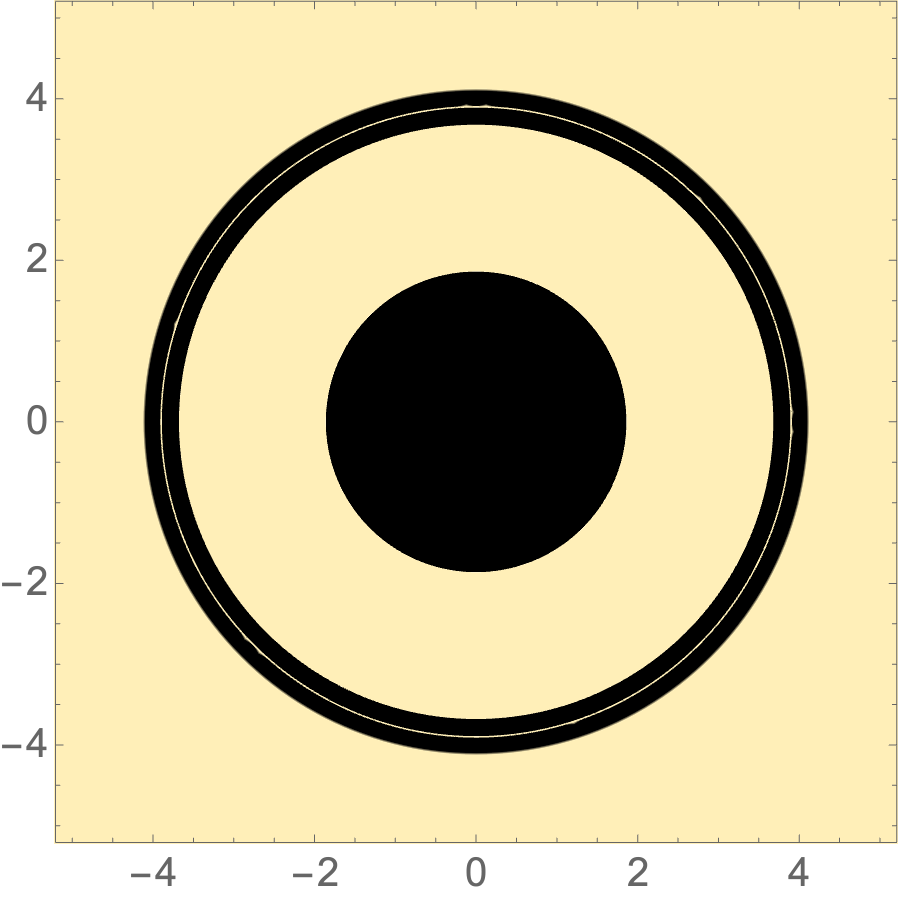} 
\caption{ \it The figure shows the backlit wormhole images. 
In the left figure the emitting screen and the observers are both in Universe II. 
In the central the emitting screen and the observers are in different Universes. 
In the right figure the emitting screens are in both Universes but the observers are in Universe II.} 
\label{back2}
\end{figure}

We now locate the observers in Universe II and perform a similar analysis as above.
First we consider that the emitting screen and the observers are both in Universe II.
The corresponding light rays are shown in Fig.~\ref{backlight2}.
The orange lines form a bright region whose edge corresponds to $b=4.1159$ in the left figure of Fig.~\ref{back2}. 
When compared to the analogous case above with both in Universe I, that is depicted in the left figure of Fig.~\ref{back1}, we note that the wormhole casts a smaller shadow in Universe II.
The red lines form a very thin bright ring inside the shadow, for which $b\in (3.9027,3.9103)$. 
Note that, as compared to the Universe I case, the bright ring is here located closer to the edge of the shadow. 
Analogous to the above discussion, there are in fact many bright rings inside the shadow that are hard to observe. 

The central plot in Fig.~\ref{back2} shows the wormhole image that is obtained when the emitting screen is in Universe I and the observers are in Universe II. 
Clearly, it is the same figure as for the case when the emitting screen in Universe II and the observers are in Universe I, i.e., the two central plots in Fig.~\ref{back1} and Fig.~\ref{back2} are identical.

Finally, when we combine the left plot and the central plot of Fig.~\ref{back2} we again obtain the right plot of Fig.~\ref{back2}.
In this case the emitting screens are in both Universes and possess the same brightness, and the observers are in Universe II.


\section{Conclusion and outlook}

In the present work, we have investigated the images of asymmetric wormholes as they appear in the different asymptotic regions. 
Of particular interest here is the presence of a light ring on only one side of the wormhole throat, which can cause light rays to be reflected by the wormhole, resulting in a typical optical/infrared/radio appearance as compared to other compact objects. 
To demonstrate this phenomenon, we have here employed the asymmetric Ellis-Bronnikov wormhole as a primary example. 
The wormhole throat, which connects Universe I and Universe II, is located at $r=-m$, and only Universe I has a light ring. 
Our results show that the behavior of null geodesics in Universe I is similar to that of null geodesics of a Schwarzschild black hole, suggesting that the wormhole appearance on this side can mimic a black hole. 
However, Universe II has no light ring, and null geodesics starting from infinity in Universe II can pass the wormhole throat into Universe I, and then be reflected back into Universe II.

Based on the analysis of the null geodesics of the Ellis-Bronnikov wormhole, we have examined the wormhole images in two scenarios. 
When using an accretion disk as the light source, we have found that the appearance of the wormhole in Universe I resembles that of a Schwarzschild black hole, with both featuring bright rings inside the shadow. 
In Universe II, the wormhole's appearance is distinctly different, however, with only bright rings visible in the dark sky. 
This unique appearance is a result of the wormhole's geometry. 
We have also considered backlit wormhole images. 
If the emitting screen lies on one side of the wormhole only, then observers who are on the same side see bright rings inside the shadow. 
Since the light ring is in Universe I, this phenomenon is clearer for observers in Universe I. 
Furthermore, since we deal with a wormhole, the emitting screen can be on the other side of the throat in Universe II. 
In this case, the backlit wormhole images are very different from those of black holes and other compact objects.

With the development of observational technology, such as Very Long Baseline Interferometry (VLBI) observations by the Event Horizon Telescope collaboration, we will obtain an increasing number of images of compact objects in the future. 
To potentially distinguish different kinds of objects, theoretical studies are essential. 
Our present study predicts new possible images describing the optical/infrared/radio appearance of traversable wormholes, which might be observable in the future. 
For further studies, one direction should be to consider images of rotating wormholes. 
A simple model for this would be the generalizations of the static Ellis-Bronnikov wormhole constructed in Ref.~\cite{Kashargin:2007mm,Kashargin:2008pk,Kleihaus:2014dla,Azad:2022qqn,Azad:2023iju}. 
This would be relevant for astrophysics, since rotating objects are more realistic. 
Another direction could be to consider a more realistic environment around the wormhole, such as a more refined model of accretion disks or free-falling matter, which might alter our results.

The code for the calculations included in this work has been made publicly available at \href{https://github.com/czhangUW/WormholeImage}{https://github.com/czhangUW/WormholeImage}.

\section*{Acknowledgments}

We thank for useful discussions with Minyong Guo at the early stage
of this work. H.H.~is grateful for support by the National Natural Science Foundation of China (NSFC) grants No.~12205123 and by the Sino-German (CSC-DAAD) Postdoc Scholarship Program, 2021 (57575640). J.K.~is grateful for support by the DFG project Ku612/18-1.

\setcounter{secnumdepth}{0} 
\section{Appendix: Determination of the deflection angle} 

In this Appendix, we briefly add some details on how we determine the curves that limit the ranges of the deflection angle.
We start from the geodesic equation (\eqref{nullgeo}) for photons, $\epsilon=0$.
The deflection angle is then calculated via
\be
\phi=\int_{\pm\infty}^{r_{end}} 1/\sqrt{\bigg(R^4(r)\left(\frac{1}{b^2}-V(r)\right)\bigg)}dr,
\ee
with the convention $\phi(\pm\infty)=0$. 
If the photons encounter a turning point at $r=r_t$, $r_{end}=r_t$. 
Otherwise the photons travel across the wormhole throat to infinity in the other Universe, thus
\be
\phi=\int_{-\infty}^{-m} 1/\sqrt{\bigg(R^4(r)\left(\frac{1}{b^2}-V(r)\right)\bigg)}dr+\int_{-m}^{+\infty} 1/\sqrt{\bigg(R^4(r)\left(\frac{1}{b^2}-V(r)\right)\bigg)}dr,
\ee
where $r=-m$ is the wormhole throat.
The location of the turning point is given by the solution of the equation
\be
\frac{1}{b^2}-V(r_t)=0.
\ee

The trajectory of a light ray with impact parameter $b$ is obtained by numerically solving \eqref{nullgeo}, which yields $r=r(\phi)$. 
If a turning point is encountered, the integration is performed only in $(\pm \infty, r_t)$ and symmetry is invoked for the trajectories in $(r_t, \pm\infty)$.
The trajectories are illustrated in polar coordinates, where we employ the physical radius $R$, 
\be
R^2=\fft{r^2+q^2-m^2}{h}
\ee
and $X=R \cos(\phi)$, $Y=R\sin(\phi)$.
Thus our terminology ``deflection angle" refers to the polar coordinate $\phi$.


We now briefly discuss how we obtain the limiting lines for the impact parameter $b$ in Figs.~\ref{ob}, \ref{wd}, and \ref{wd2}, starting with Fig.~\ref{ob}.
In this case the light source, the accretion disk, and the observers are in Universe I.
The purple line denotes the deflection angle $\phi_{purple}$ for a light ray from infinity of Universe I to its first intersection of the ISCO, $r^{1}_{isco}$,  
\be
\phi_{purple}=\int_{-\infty}^{r^{1}_{isco}}
1/\sqrt{\bigg(R^4(r)\left(\frac{1}{b^2}-V(r)\right)\bigg)}dr .
\ee
It delimits the orange light rays in the left figure of Fig.~\ref{ob}.
The critical value is obtained when a light ray hits the accretion disk for the first time and $\phi_{purple}=-\ft{\pi}{2}$. 
This happens for $b=5.2614$. 
Thus the bright region outside the black hole shadow is formed by light rays with impact parameter $b$ in the range $b\in (5.2614, +\infty)$.
A part of this range is indicated by the upper blue line in Fig.~\ref{ob}.
It is associated with transfer function $r_n(b)$ with $n=1$.

The red line denotes the deflection angle $\phi_{red1}$ for a light ray from infinity in Universe I to its second intersection of the ISCO, $r^2_{isco}$,
\be
\phi_{red1}=
\int_{-\infty}^{r^{2}_{isco}}
1/\sqrt{\bigg(R^4(r)\left(\frac{1}{b^2}-V(r)\right)\bigg)}dr .
\ee
The black line denotes the deflection angle $\phi_{black1}$ for a light ray from infinity to a turning point $r_0$ and back to infinity in the same Universe I,
\be
\phi_{black1}=2\phi(r_0)
=2\int_{-\infty}^{r_0}
1/\sqrt{\bigg(R^4(r)\left(\frac{1}{b^2}-V(r)\right)\bigg)}dr .
\ee
Thus the deflection angle $\phi_{black1}$ is twice the deflection angle of a light ray from infinity to the turning point in Universe I.

The light ray with $\phi_{black1}=-\ft{3\pi}{2}$ intersects the accretion disk at infinity.
It is one of the critical light rays, and has $b=4.4097$. 
The light ray with $\phi_{red1}=-\ft{3\pi}{2}$ intersects the accretion disk at $(R=R_{isco}, \phi=-\ft{3\pi}{2})$ and is another critical light ray.
It has $b=4.0593$.
Thus the second blue line in Fig.~\ref{ob} gives rise to the first bright ring of the wormhole images in Fig.~\ref{image1}, for which $n=2$.
Analogously, the light rays with $\phi_{black1}=-\ft{5\pi}{2}, -\ft{7\pi}{2},\cdots$ also form critical light rays, and likewise light rays with $\phi_{red1}=-\ft{5\pi}{2}, -\ft{7\pi}{2},\cdots$.
Consequently, the second bright ring in Fig.~\ref{image1} corresponds to $b\in (3.9087, 3.9189)$, corresponding to $n=3$, that cannot be observed in the image, Fig.~\ref{image1}.

Next we discuss the case with the light source, the accretion disk, in Universe I and the observers in Universe II, illustrated in Fig.~\ref{wd}. 
In this case, only light rays travelling across the wormhole throat to Universe I can hit the accretion disk. 
Such light rays require $b<b_c$. 
Considering a specific light ray with $b<b_c$, its deflection angle when arriving at the throat is given by 
\be
\phi_t=\int_{+\infty}^{-m} 1/\sqrt{\bigg(R^4(r)\left(\frac{1}{b^2}-V(r)\right)\bigg)}dr .
\ee
This light ray will continue from the throat to the ISCO corresponding to
the deflection angle 
\be
\phi_{isco}=\int_{-m}^{r_{isco}} 1/\sqrt{\bigg(R^4(r)\left(\frac{1}{b^2}-V(r)\right)\bigg)}dr .
\ee
Thus the deflection angle from infinity in Universe II to the ISCO in Universe I is given by
\be
\phi_{red2}=\phi_t+\phi_{isco}=\int_{+\infty}^{r_{isco}} 1/\sqrt{\bigg(R^4(r)\left(\frac{1}{b^2}-V(r)\right)\bigg)}dr ,
\ee
and is illustrated by the red line in Fig.~\ref{wd}.
On the other hand, when a light ray will keep moving to infinity in Universe I, the total deflection angle for this light ray from infinity in Universe II to infinity in Universe I is given by
\be\label{blackline}
\phi_{black2}=\int_{+\infty}^{-\infty} 1/\sqrt{\bigg(R^4(r)\left(\frac{1}{b^2}-V(r)\right)\bigg)}dr ,
\ee
and is illustrated by the black line in Fig.~\ref{wd}.
The function $\phi_{black2}=\phi_{black2}(b)$ exists for any $b<b_c$. 

The light ray with $\phi_{red2}=\ft{\pi}{2}$ passes the point $(R=R_{isco}, \phi=\ft{\pi}{2})$ and is one of the critical light rays. 
It has $b=3.8052$.
The light ray with $\phi_{black2}=\ft{\pi}{2}$ intersects the accretion disk at infinity and is another critical light ray.
It has $b=3.6799$. 
The first blue line in Fig.~\ref{wd} ranges between these two critical values of $b$ and gives rise to the first bring ring of the wormhole images in Fig.~\ref{image2}, corresponding to $n=1$, since
these light rays hit the accretion disk directly.
Analogously, the light rays from $\phi_{red2}=\ft{3\pi}{2}$ ($b=2.3679$) to $\phi_{black2}=\ft{3\pi}{2}$ ($b=1.8614$) form the range of the second bright ring, 
shown by the shorter blue line in Fig.~\ref{wd}, corresponding to $n=2$.
The third bright ring is only seen in the intensity plot of Fig.~\ref{image2}, but would be hard to observe in the image.

Figure \ref{wd2} considers that the observers and the emitting screen are located in different Universes. 
Without loss of generality, we assume that the observers are at infinity in Universe II and the emitting screen is at infinity in Universe I. 
The black line in the Fig.~\ref{wd2} denotes the deflection angle for this process. 
We obtained this deflection angle already when studying Fig.~\ref{wd} (\eqref{blackline}).
The only difference with respect to the case above is that the light source is now the emitting screen, and not the accretion disk. 
Therefore the boundary of the bright regions is determined by the condition that the light rays can hit the emitting screen at infinity. 
Such light rays have their deflection angle in the ranges $\phi=(\ft{\pi}{2}, \ft{3\pi}{2})$, $\phi=(\ft{5\pi}{2}, \ft{7\pi}{2})$ and so on. 
The two blue lines in Fig.~\ref{wd2} denote the ranges of $b$ in $\phi=(\ft{\pi}{2}, \ft{3\pi}{2})$ and $\phi=(\ft{5\pi}{2}, \ft{7\pi}{2})$.

\newpage

\bibliography{ref}
\bibliographystyle{utphys}

\end{document}